\documentclass[nofootinbib, superscriptaddress]{revtex4-2}
\usepackage[utf8]{inputenc}
\usepackage{bm, amsthm, amsmath, amsfonts, amssymb, color, graphicx, natbib}
\usepackage{slashed}
\usepackage{xcolor}
\usepackage{verbatim}
\usepackage{float}
\usepackage{gensymb}
\usepackage{graphicx}   
\usepackage{subcaption}
\usepackage{caption}     
\graphicspath{{fig/}}
\usepackage{hyperref}
\usepackage[autostyle, english = american]{csquotes}
\MakeOuterQuote{"}
\hypersetup{
	colorlinks=true,
	linkcolor=blue,
	filecolor=blue,      
	urlcolor=blue,
	citecolor=blue
}

\newcommand{\dotdeg}{\rlap{.}^\circ}

\begin{document}

\title{Reconciling cosmic dipolar tensions with a gigaparsec void} 

\author{Tingqi Cai}
\email{tcaiac@connect.ust.hk}

\affiliation{Department of Physics, The Hong Kong University of Science and Technology,\\
Clear Water Bay, Kowloon, Hong Kong, P.R.China}
\affiliation{Jockey Club Institute for Advanced Study, The Hong Kong University of Science and Technology,\\
Clear Water Bay, Kowloon, Hong Kong, P.R.China}

\author{Qianhang Ding}
\email{qdingab@connect.ust.hk}

\affiliation{Department of Physics, The Hong Kong University of Science and Technology,\\
Clear Water Bay, Kowloon, Hong Kong, P.R.China}
\affiliation{Jockey Club Institute for Advanced Study, The Hong Kong University of Science and Technology,\\
Clear Water Bay, Kowloon, Hong Kong, P.R.China}
\affiliation{Cosmology, Gravity and Astroparticle Physics Group, Center for Theoretical Physics of the Universe,
Institute for Basic Science (IBS), Daejeon, 34126, Korea}

\author{Yi Wang}
\email{phyw@ust.hk}

\affiliation{Department of Physics, The Hong Kong University of Science and Technology,\\
Clear Water Bay, Kowloon, Hong Kong, P.R.China}
\affiliation{Jockey Club Institute for Advanced Study, The Hong Kong University of Science and Technology,\\
Clear Water Bay, Kowloon, Hong Kong, P.R.China}

\begin{abstract}
  Recent observations indicate a $4.9\sigma$ tension between the CMB and quasar dipoles. This tension challenges the cosmological principle. We propose that if we live in a gigaparsec scale void, the CMB and quasar dipolar tension can be reconciled. This is because we are unlikely to live at the center of the void. And a within 10\% offset from the center will impact the quasars and CMB differently in their dipolar anisotropies. As we consider a large and thick void, our setup can also ease the Hubble tension.
\end{abstract}

\maketitle

\section{Introduction}

The cosmological principle is a fundamental postulate in modern cosmology, assuming that the universe is homogeneous and isotropic on large scales, independent of the location of observers \cite{1937RSPSA.158..324M}. Based on the cosmological principle, the Lambda cold dark matter ($\Lambda \mathrm{CDM}$) model \cite{1970ApJ...159..379R, PhysRevLett.52.2090, 1985Natur.317..595F} is established and considered as the standard model of cosmology. 

With the development of precision cosmology, new observable results show hints of inconsistencies between observations and the cosmological principle, e.g., the detection of cosmic dipole such as cosmic microwave background (CMB) dipole \cite{Planck:2018nkj} and quasar dipole \cite{Secrest:2020has} may be inconsistent and indicate an anisotropic observable universe.
Also, the Hubble tension between its local measurements \cite{Riess:2016jrr, Brout:2022vxf} and the value from Planck \cite{Planck:2016kqe}, which may be interpreted as the existence of a local inhomogeneous structure \cite{Ding:2019mmw, Haslbauer:2020xaa}.
A fundamental explanation is needed for these problems.

In reconciling the inconsistency between observations and the cosmological principle, a number of models have been proposed, such as introducing new physical scenarios in solving the Hubble tension \cite{Poulin:2018cxd, Bernal:2016gxb, Agrawal:2019lmo, Lin:2019qug, Kenworthy:2019qwq, Vagnozzi:2019ezj, Fung:2021fcj, Camarena:2022iae} and using the peculiar motion of observers in explaining the cosmic dipole \cite{1984MNRAS.206..377E, Lavaux:2008th}. However, the tension between the CMB dipole and quasar dipole remains unsolved. The amplitude of the quasar dipole is over twice as large as the expected value in the kinematic interpretation of the CMB dipole. In the same analysis, the inconsistency is reported to be as significant as $4.9 \sigma$ \cite{Secrest:2020has}. Such a large quasar dipole anisotropy still exists after removing the standard kinematic dipole in the CMB frame, which may indicate the existence of a local anisotropic structure \cite{Secrest:2022uvx}. Meanwhile, some theoretical attempts have been done in solving this dipolar tension, see \cite{Dalang:2021ruy, Domenech:2022mvt, Krishnan:2022qbv} for details.

The introduction of a Gpc-scale local void changes the story. A local underdense region with us as observers inside could cause additional peculiar motion of nearby supernovae (SNe). A Gpc-scale void makes sure that almost all detected SNe live inside, which biases local measurement of the Hubble parameter and eases the Hubble tension (see \cite{Ding:2019mmw} for more details). Since it is very unlikely for the position of the Milky Way to be located at the exact center of this local void, the observational anisotropy would be induced by an off-center location of observers inside the void \cite{Alnes:2006pf, Alnes:2006uk, Nistane:2019yzd}. 

In this article, we propose that the amplitude inconsistency between CMB and quasar dipoles can also be explained by our offset from the center of a void. Due to the existence of such an anisotropic spacetime in an off-center void, the detected photons from different directions have experienced different cosmic expansion histories, which causes an anisotropy in their cosmic redshifts. The CMB dipole that measured from CMB temperature perturbations can be in part attributed to the redshift dipole anisotropy. Meanwhile, quasar dipole that measured from hemisphere quasar number counting \cite{Rameez:2017euv}, can be partly attributed to the anisotropic matter distribution in the view of an off-center observer, where the void is as large as the average distance from us to distant quasars. Depend on void profile, the observed amplitude of dipole structures at different redshifts varies, especially has a peak amplitude around the void boundary. Such a peak feature around void boundary would provide an opportunity for us to explain the observed dipolar tension between CMB dipole and quasar dipole. In addition, Another part of origin of observed dipoles comes from the peculiar velocity of our solar system in the rest frame of CMB. To fully understand the dipolar tension in local void scenario, a combination study on void-induced dipole and peculiar velocity induced dipole would be detailed investigated in this work.

\section{Origin of a Gpc-scale void}

In standard cosmological scenarios where structures originate from a Gaussian random density field, the presence of a Gpc-scale void is unlikely, due to $\sigma_8 \simeq 0.81$ \cite{Planck:2018vyg}, which shows that the amplitude of fluctuations is statistically suppressed on comoving scales much larger than $10 \, \mathrm{Mpc}$. Therefore, a Gpc-scale void, if it exists, deserves a distinct primordial origin. Multi-stream inflation \cite{Li:2009sp} is a potential mechanism to produce a Gpc-scale void. Multi-stream inflation can also generate position space features in cosmology, such as multiverse structures \cite{Li:2009me}, CMB cold spots \cite{Afshordi:2010wn}, initial primordial black holes clustering \cite{Ding:2019tjk}, and primordial stellar bubbles \cite{Cai:2021zxo}.

\begin{figure}[ht]
	\centering
	\includegraphics[width=0.8\textwidth]{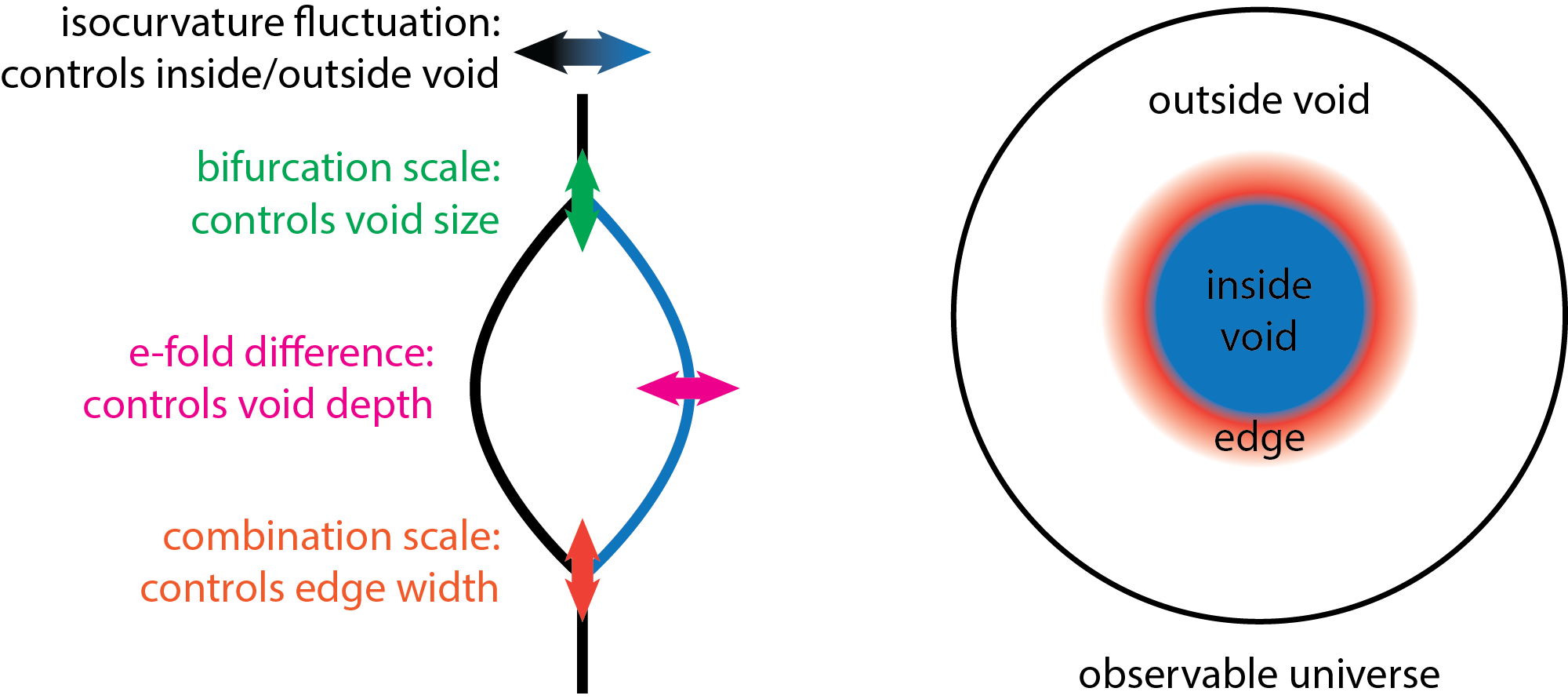}
	\caption{This figure demonstrates how a cosmic void is generated from multi-stream inflation, which is taken from the Fig.~1 in \cite{Ding:2019mmw}. \textit{Left panel:} The multi-stream inflation in the primordial universe, where inflationary trajectory encounters a barrier, and bifurcates into two paths with different probabilities. With a larger probability inflaton rolls into the black trajectory, while a smaller probability it rolls into the blue trajectory, which has a larger e-folding number and becomes a void in late universe. \textit{Right panel:} The late-time universe with a void structure embedding arised from the multi-stream inflation.}
	\label{fig:multi_stream}
\end{figure}

In multi-stream inflation, the inflationary trajectory may encounter a barrier and then bifurcate into two paths, which experience different inflationary potentials. The inflationary dynamics and the void profile are related as follows: (i) The density contrast between inside and outside the void is determined by the e-folding number difference between the two trajectories. The trajectory which is now a void has fewer e-folds of inflation, by $\delta \rho/\rho \sim \delta N$, where $\delta N$ is the e-folding number difference. (ii) The size of the void is determined by the comoving scale during inflation when bifurcation happened. Thus, the size of the void is a free parameter which can be made Gpc. (iii) The thickness of the boundary between inside and outside the void is determined by the combination scale of the two trajectories. Thus, the void originated from multi-stream inflation can have a smooth profile, which is very important for our model to be consistent with kSZ constraints, as we will emphasize later. An illustration is shown in Fig.~\ref{fig:multi_stream}. 

\section{Void cosmology}

In the universe with a local void embedded, the spacetime is no longer homogeneous and isotropic; therefore, the FRW metric cannot be applicable for void cosmology. In order to describe the spacetime in void cosmology, we start from a simple scenario; the shape of the void is spherical. In the view of the observer at the void center, the spacetime can be described by an inhomogeneous and isotropic metric, which is the well-known Lemaitre-Tolman-Bondi (LTB) metric \cite{Lemaitre:1933gd, Tolman:1934za, Bondi:1947fta} as follows,
\begin{equation}\label{eq:LTB_metric}
    \mathrm{d}s^2=c^2\mathrm{d}t^2 -\frac{R'(r,t)^2}{1-k(r)}\mathrm{d}r^2-R^2(r,t)\mathrm{d}\Omega^2~,
\end{equation}
where $R'(r,t) = \partial R(r,t)/\partial r$. In a homogeneous scenario, the spacetime metric can be recovered from the LTB metric to the FRW metric via $R(r,t) = a(t) r$, $k(r) = k r^2$, where $a(t)$ is the usual scale factor. We follow \cite{Garcia-Bellido:2008vdn} to apply the LTB metric in the Einstein equation to obtain the Friedmann equation in this inhomogeneous and isotropic spacetime as follows
\begin{equation}\label{eq:Hubble}
	H(r,t)^2 \equiv \frac{\dot{R}(r,t)^2}{R(r,t)^2}=H_0(r)^2 \left(\Omega_M(r) \frac{R_0(r)^3}{R(r,t)^3} + \Omega_k(r)\frac{R_0(r)^2}{R(r,t)^2}+\Omega_\Lambda(r)\right)~,
\end{equation}
where $H_0(r)$ is the space-dependent local Hubble parameter at present cosmic time defined as $H(r,t_0)$ and $R_0(r)$ is defined as $R_0(r) \equiv R(r,t_0)$. $\Omega_X(r)\equiv\rho_X(r)/\rho_c(r)$ with $X = M, \, k, \, \Lambda$, representing matter, curvature, and dark energy, respectively. The critical energy density is $\rho_c(r)\equiv3H_0(r)^2/8\pi G$. The density parameters satisfy $\Omega_M(r)+\Omega_k(r)+\Omega_\Lambda(r)=1$.

\begin{figure}[ht]
	\centering
	\includegraphics[width=0.46\textwidth]{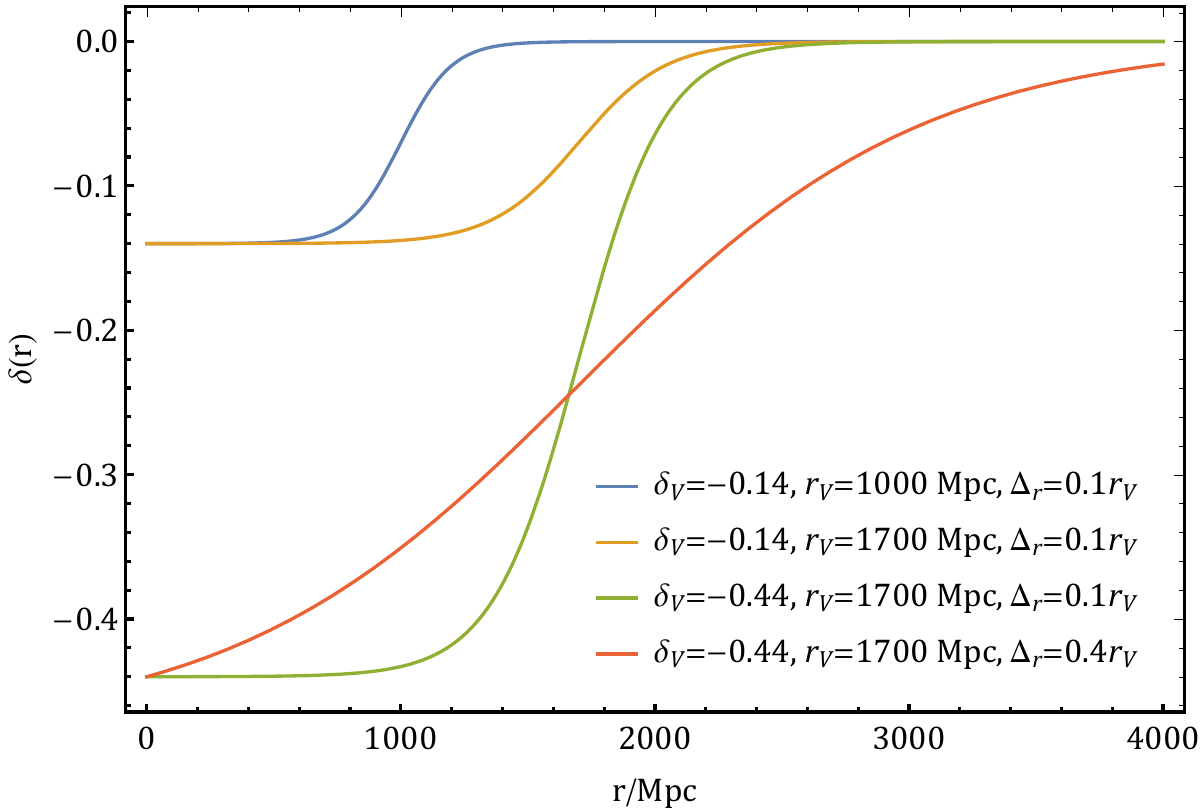}
	\caption{Some example of void profiles, void depth $\delta_V$ determines the matter density contrast, void radius $r_V$ determines the size of low matter density region inside the void, and the width of the void boundary $\Delta_r$ determines the thickness of void that transforms from low matter density to high matter density.}
	\label{fig:void_profile}
\end{figure}

In order to describe the void spacetime, the matter density contrast of a void is defined as follows
\begin{equation}
	\delta \equiv \frac{\rho_M(r) - \rho_M(\infty)}{\rho_M(\infty)}~.
\end{equation}
We follow \cite{Kenworthy:2019qwq} to parameterize the void profile as
\begin{equation}\label{eq:void_profile}
	\delta(r) = \delta_V\frac{1-\tanh((r-r_V)/2\Delta_r)}{1+\tanh(r_V/2\Delta_r)}~,
\end{equation}
where $\delta_V$ and $r_V$ are the depth and radius of the void, and $\Delta_r$ is the width of the void boundary. Some examples of void profile are shown in Fig.~\ref{fig:void_profile}, 

\begin{figure}[ht]
	\centering
	\includegraphics[width=0.46\textwidth]{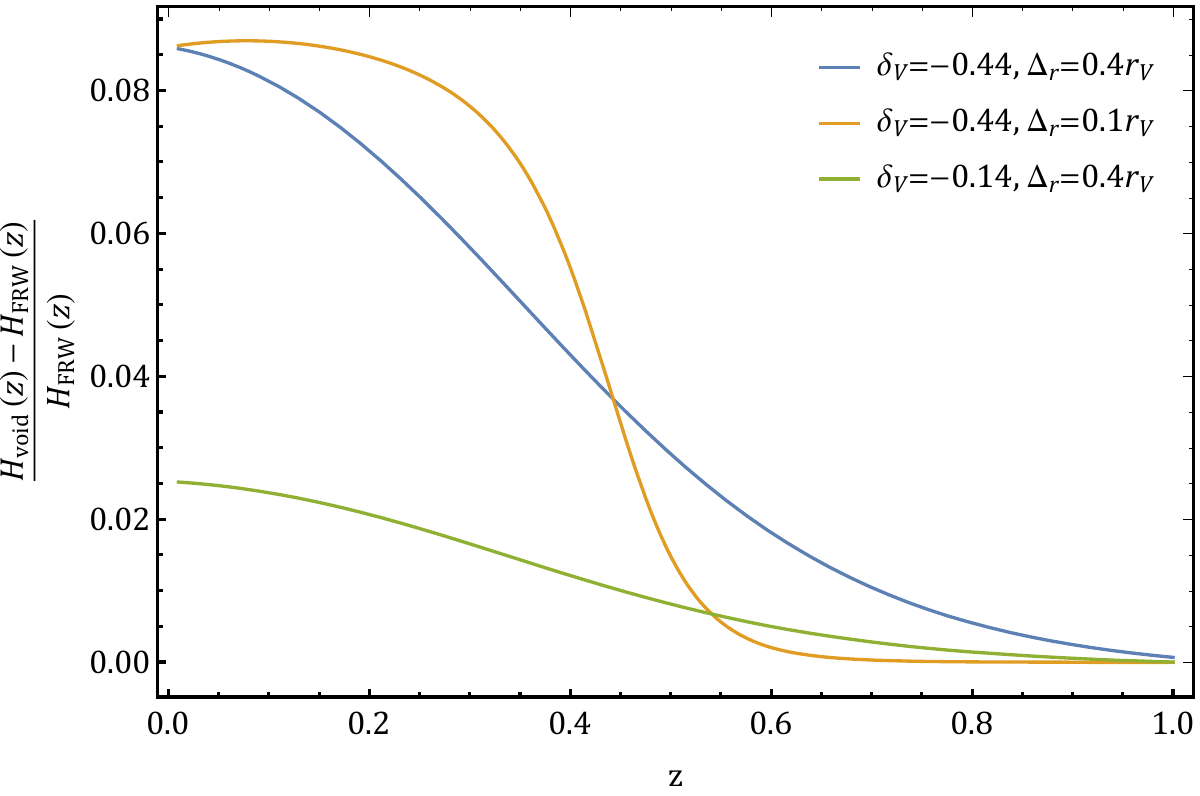}
	\caption{The difference of Hubble parameter at different redshifts between various void cosmologies and FRW cosmology. $\delta_V$ is the depth of the void, $\Delta_r$ is the thickness of the void, and the radius of void $r_V$ is set as $1.7 \, \mathrm{Gpc}$. The FRW parameter we calculate is $H_0 = 67.4 \, \mathrm{km} \, \mathrm{s}^{-1} \, \mathrm{Mpc}^{-1}$, $\Omega_M = 0.315$, and $\Omega_\Lambda = 0.685$. }
	\label{fig:Hubble}
\end{figure}

We follow \cite{Kenworthy:2019qwq} to use matter density contrast and current Hubble parameter to express the energy density as
\begin{align}\label{eq:rho_m}
	\rho_M(r) & \propto \Omega_M(r) H_0(r)^2 = \Omega_{M,\mathrm{out}} (1+\delta(r)) H_{0,\mathrm{out}}^2~,\\\label{eq:rho_lambda}
	\rho_\Lambda(r) & \propto \Omega_\Lambda(r) H_0(r)^2 = (1-\Omega_{M,\mathrm{out}}) H_{0,\mathrm{out}}^2 = \mathrm{const}~,\\\label{eq:rho_k}
	\rho_k(r) & \propto \Omega_k(r) H_0(r)^2 = (1 - \Omega_M(r) - \Omega_\Lambda(r)) H_0(r)^2 = H_0(r)^2 - (1 + \Omega_{M,\mathrm{out}} \delta(r)) H_{0,\mathrm{out}}^2~.
\end{align}
Here and hereafter the subscripts ‘‘out'' denote quantities outside the void. Apply Eqs.~(\ref{eq:rho_lambda} -- \ref{eq:rho_k}) in Eq.~\eqref{eq:Hubble}, it can be expanded as follows,
\begin{equation}
	\frac{\dot{R}(r,t)^2}{R(r,t)^2} = \Omega_{M,\mathrm{out}} (1+\delta(r)) H_{0,\mathrm{out}}^2 \frac{R_0(r)^3}{R(r,t)^3} + (H_0(r)^2 - (1+\Omega_{M,\mathrm{out}}\delta(r)) H_{0,\mathrm{out}}^2) \frac{R_0(r)^2}{R(r,t)^2} + (1-\Omega_{M,\mathrm{out}}) H_{0,\mathrm{out}}^2~.
\end{equation}
To fix the gauge, we fix $R(r,t_0) = R_0(r) = r$, which gives
\begin{equation}\label{eq:R}
	\frac{\dot{R}(r,t)^2}{R(r,t)^2} = \Omega_{M,\mathrm{out}} (1+\delta(r)) H_{0,\mathrm{out}}^2 \frac{r^3}{R(r,t)^3} + (H_0(r)^2 - (1+\Omega_{M,\mathrm{out}}\delta(r)) H_{0,\mathrm{out}}^2) \frac{r^2}{R(r,t)^2} + (1-\Omega_{M,\mathrm{out}}) H_{0,\mathrm{out}}^2~.
\end{equation}
Then we integrate Eq.~\eqref{eq:R}, it gives
\begin{equation}\label{eq:tB}
	t_B(r) = \int_0^r dR \, R^{-1} \left[\Omega_M(r) H_0(r)^2 \left(\frac{r}{R}\right)^3 + \Omega_k(r) H_0(r)^2 \left(\frac{r}{R}\right)^2 + \Omega_\Lambda(r) H_0(r)^2 \right]~.
\end{equation}
Here, $t_B$ is the cosmic time since the big bang. Following \cite{Kenworthy:2019qwq}, we set $t_B(r) = t_B = \text{const}$ and $t_0 = t_B$. In the region outside the void $r \gg r_V$, we have $\delta(r) = 0$ and $H_0(r) = H_{0,\mathrm{out}}$, then Eq.~\eqref{eq:tB} can be simplified as follows,
\begin{align}\label{eq:bigbangtime}
	t_B = \int_0^1 \frac{d a_\mathrm{out}}{H_{0,\mathrm{out}} [\Omega_{M,\mathrm{out}} a_\mathrm{out}^{-1} + \Omega_{\Lambda, \mathrm{out}} a_\mathrm{out}^2]^{1/2}}~.
\end{align}
Here, $a_\mathrm{out}$ is the scale factor outside the void. Given a set of $(H_{0, \mathrm{out}}, \Omega_{M, \mathrm{out}}, \Omega_{\Lambda, \mathrm{out}})$, Eq.~\eqref{eq:bigbangtime} can determine a fixed $t_B$, which can be applied in Eq.\eqref{eq:tB} to obtain $H_0(r)$. Then $R(r,t)$ can be numerically solved from the following equation,
\begin{align}\label{eq:solveR}
	\frac{\partial R}{\partial t} = R \left[ H_0(r)^2 \Omega_M(r) \frac{r^3}{R^3} + H_0(r)^2 \Omega_k(r) \frac{r^2}{R^2} +H_0(r)^2 \Omega_\Lambda(r) \right]^{1/2}~.
\end{align}
In Eq.~\eqref{eq:solveR}, we set the initial condition as $R(r,t)=r$ and solve this differential equation backward in time from $t_B$ to $t (< t_B)$ to obtain $R(r,t)$. The dependence of $H(r,t)$ and $R(r,t)$ on redshift can be found by solving the null geodesic equation in void spacetime as follows,
\begin{equation}
	\frac{dt}{dr} = -\frac{1}{c} \frac{R'(r,t)}{\sqrt{1-k(r)}}~,\quad \frac{1}{1+z} \frac{dz}{dr} = \frac{1}{c} \frac{\dot{R}'(r,t)}{\sqrt{1-k(r)}}~.
\end{equation}
With the solution of the geodesic equation, we can calculate the behavior of cosmological parameters in various void cosmologies and compare them with the parameters in the FRW universe, such as the Hubble parameter, which can be expressed as $H_\mathrm{FRW}(z) = H_{0,\mathrm{out}} \sqrt{\Omega_M(1+z)^3 + \Omega_\Lambda}$. The result is shown in Fig.~\ref{fig:Hubble},
We can find that different void parameters control various behaviors of the local universe. A deeper void can effectively increase the value of the local Hubble parameter and decrease the Hubble tension. A larger void can affect the local cosmological parameters up to higher redshifts, which can help explain the difference of cosmological parameters between local measurements and high redshift measurements. A smoother void can make the behavior of cosmological parameters evolve smoothly, which could help avoid some observational constraints such as constraints from kSZ effect (for more details, see Sec.~\ref{subsec:kSZ}).

\section{Observational constraints}
The existence of such a Gpc-scale void affects different observations of e.g. Type Ia SNe, baryon acoustic oscillations (BAO) and CMB. The cosmic influence of a void with a suitable profile should be consistent with these observational constraints. In the following section, we briefly review results in \cite{Ding:2019mmw} about how observations constrain such a Gpc-scale void (see also \cite{Biswas:2010xm, Camarena:2021mjr} and references therein).

\subsection{Type Ia supernovae}
Type Ia SNe provide a key measurement for the local universe. From their light curves at redshift $0 < z < 2.3$ \cite{Scolnic:2021amr}, the luminosity distance-redshift relation can be determined, which gives a local Hubble parameter $H_0 = 73.3 \pm 1.1 \mathrm{km} \, \mathrm{s}^{-1} \, \mathrm{Mpc}^{-1}$ in $\Lambda \mathrm{CDM}$ cosmology \cite{Brout:2022vxf}, while Planck gives Hubble parameter $H_0 = 66.9 \pm 0.6 \mathrm{km} \, \mathrm{s}^{-1} \, \mathrm{Mpc}^{-1}$ \cite{Planck:2016kqe}. Such a tension in the Hubble parameter could be eased by a local Gpc-scale void with the Milky Way inside, which causes nearby SNe to live inside the void and have significant positive peculiar velocities, biasing the local measurement of the Hubble parameter. 

\begin{figure}[ht]
	\centering
	\includegraphics[width=0.46\textwidth]{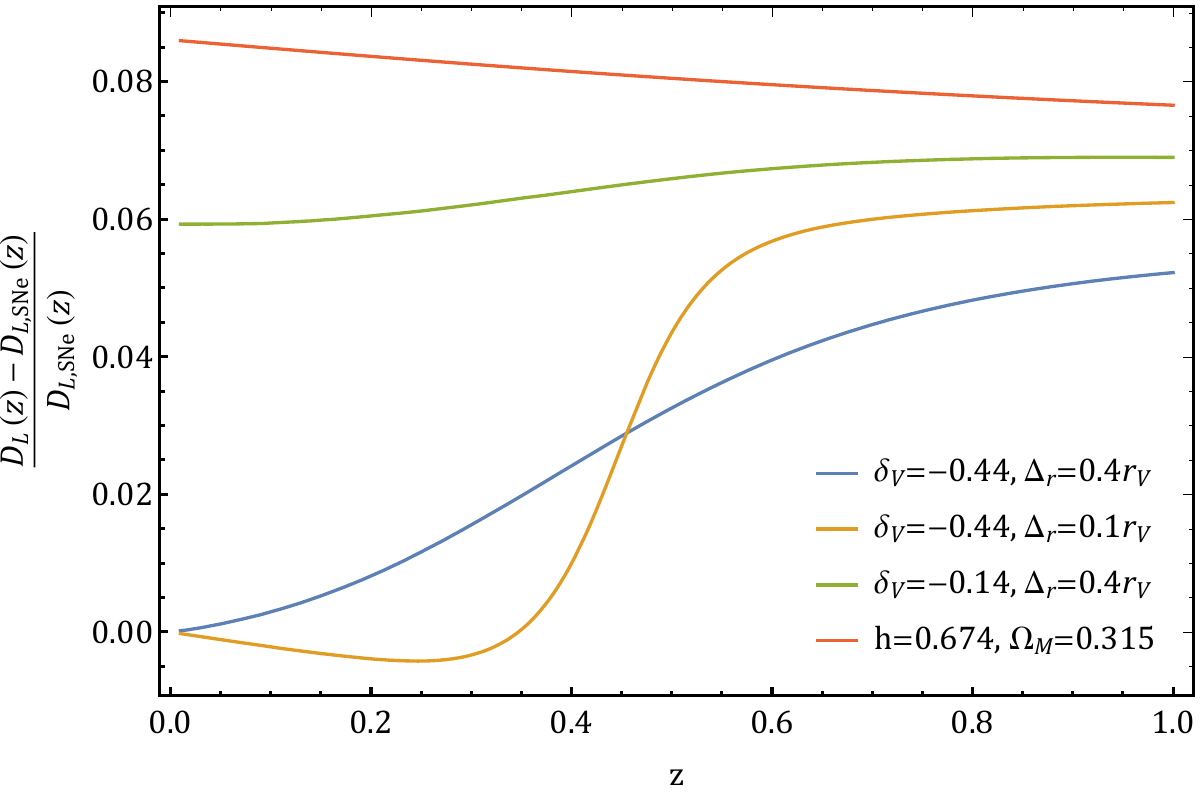}
	\caption{The difference of luminosity distance--redshift relation between various void cosmologies and FRW cosmology with parameter $(h, \Omega_M, \Omega_\Lambda) = (0.732, 0.3, 0.7)$ from local Type Ia measurement \cite{Riess:2016jrr}, where $h \equiv H_0/100 \, \mathrm{km} \, \mathrm{s}^{-1} \, \mathrm{Mpc}^{-1}$. $\delta_V$ is the depth of the void, $\Delta_r$ is the thickness of the void, and the radius of void $r_V$ is set as $1.7 \, \mathrm{Gpc}$. The FRW cosmology with parameter $(h, \Omega_M, \Omega_\Lambda) = (0.674, 0.315, 0.685)$ from Planck 2018 \cite{Planck:2018vyg} is also compared.}
	\label{fig:lumin_dis}
\end{figure}

As shown in Fig.~\ref{fig:Hubble}, the behavior of the Hubble parameter inside the local void can help ease Hubble tension. And the profile of the void determines the behavior of the Hubble parameter in void spacetime. Especially, when we consider the size of the void is up to Gpc scale, nearby Type Ia SNe are located inside this void and their observation results could be explained by such a local void, such as luminosity distance-redshift relation, $D_L(z)$, which is expressed in void cosmology as follows,
\begin{align}
	D_{L,\mathrm{void}}(z) = (1+z)^2 R(r(z), t(z))~.
\end{align}
Meanwhile, luminosity distance in FRW cosmology is $D_L(z) = (1+z)\int_0^z c/H(z') dz'$. Their local behaviors are shown in Fig.~\ref{fig:lumin_dis}.
We can find the local behavior of void cosmology with void depth $\delta_V = -0.44$, can well fit local luminosity distance--redshift relation up to $z \sim 0.4$, which shows void cosmology with a large void depth can effectively describe the behavior of the local universe compared with FRW cosmology from Planck 2018 in Fig.~\ref{fig:lumin_dis}. Also, the Hubble tension can be solved in such a deeper void with void depth $\delta_V = -0.44$ as shown in Fig.~\ref{fig:Hubble}; however, some other observations put strong constraints on void depth, see Sec.~\ref{subsec:BAO} and \ref{subsec:kSZ} for more details.

\subsection{Baryon acoustic oscillations}\label{subsec:BAO}
BAO scale measurements at different redshifts \cite{2011MNRAS.416.3017B, Ross:2014qpa, BOSS:2016wmc, Planck:2018vyg} provide standard rulers to constrain cosmological models. The observable used from BAO measurement is $(\Delta \theta^2 \Delta z)^{1/3}$, which is model independent and can test various cosmologies. In FRW cosmology, it can be expressed as
\begin{align}\label{eq:frw_bao}
	(\Delta \theta^2 \Delta z)^{1/3} = \frac{z_\mathrm{BAO}^{1/3} r_d}{D_V^\mathrm{FRW}(z_\mathrm{BAO})}~,
\end{align}
where $r_d$ is the sound horizon at the drag epoch, and $D_V^\mathrm{FRW}$ can be expressed in the flat FRW case,
\begin{align}
	D_V^\mathrm{FRW}(z_\mathrm{BAO}) = \frac{1}{H_0} \left[\frac{z_\mathrm{BAO}}{E(z_\mathrm{BAO})} \left(\int_0^{z_\mathrm{BAO}} \frac{dz}{E(z)}\right)^2\right]^{1/3}~,
\end{align}
where $E(z) \equiv H(z)/H_0$. Meanwhile, we follow \cite{Biswas:2010xm} to express $(\Delta \theta^2 \Delta z)^{1/3}$ in void cosmology as,
\begin{align}\label{eq:void_bao}
	 (\Delta \theta^2 \Delta z)^{1/3} = \left[\frac{(1+z_\mathrm{BAO}) \dot{R}'_\mathrm{BAO}}{R'(r_\mathrm{BAO}, t_d(r_\mathrm{BAO})) R^2(r_\mathrm{BAO}, t_d(r_\mathrm{BAO}))}\right]^{1/3} \frac{r_d(r_\mathrm{BAO})}{1+z_d(r_\mathrm{BAO})}~.
\end{align}
Here, $r_\mathrm{BAO}$, $t_\mathrm{BAO}$ are radius and time corresponding with $z_\mathrm{BAO}$, and $R_\mathrm{BAO}$ is defined as $R(r_\mathrm{BAO}, t_\mathrm{BAO})$. $r_d(r)$ and $z_d(r)$ are $r$-dependent sound horizon and redshift at the drag epoch, respectively.

\begin{figure}[ht]
	\centering
	\includegraphics[width=0.46\textwidth]{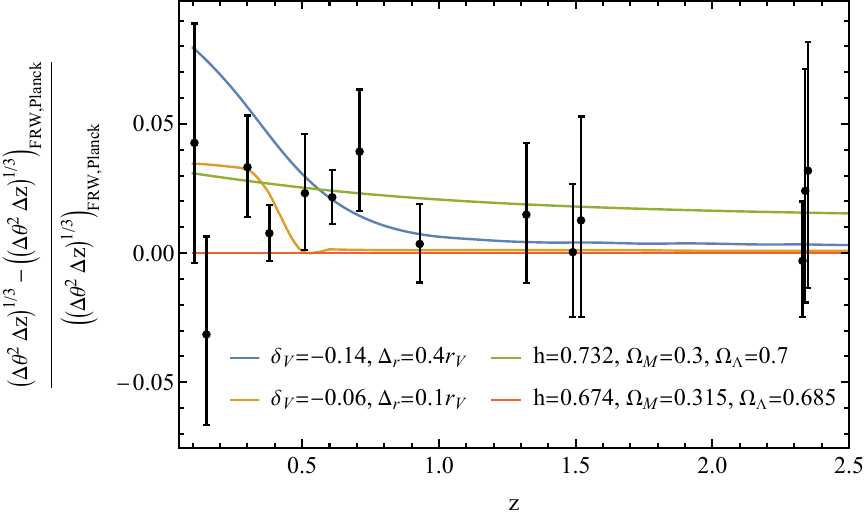}
	\caption{The difference of $(\Delta \theta^2 \Delta z)^{1/3}$ between various void cosmologies and FRW cosmology with parameter $(h, \Omega_M, \Omega_\Lambda) = (0.674, 0.315, 0.685)$ from Planck 2018 \cite{Planck:2018vyg}. $\delta_V$ is the depth of the void, $\Delta_r$ is the thickness of the void, and the radius of void $r_V$ is set as $1.7 \, \mathrm{Gpc}$.  The FRW cosmology with parameter $(h, \Omega_M, \Omega_\Lambda) = (0.732, 0.3, 0.7)$ from local Type Ia measurement \cite{Riess:2016jrr} is also compared. BAO data for $(\Delta \theta^2 \Delta z)^{1/3}$ is taken from various observations \cite{beutler20116df, Ross:2014qpa, BOSS:2016wmc, Blomqvist:2019rah, deSainteAgathe:2019voe, Ata:2017dya, DESI:2024mwx}.}
	\label{fig:BAO}
\end{figure}

Based on Eqs.~\eqref{eq:frw_bao} and \eqref{eq:void_bao}, we can use the observational BAO data to test various FRW and void cosmologies as shown in Fig.~\ref{fig:BAO}.
We can find that the redshift dependence of spatial curvature in the local void biases a lager value of $(\Delta \theta^2 \Delta z)^{1/3}$ and help fit the data better, which is also mentioned in \cite{Biswas:2010xm}. Also, we can notice that the low matter density at $z<1$ is favored by BAO data and the void depth is constrained by low redshift measurement of $(\Delta \theta^2 \Delta z)^{1/3}$, a void with depth $\delta_V < -0.14$ is disfavored.

\subsection{Kinetic Sunyaev-Zel'dovich effect} \label{subsec:kSZ}
A local void could cause the temperature perturbation at small scales of CMB, due to interactions between bulk flow electrons and CMB photons, which is called the kinetic Sunyaev-Zel'dovich (kSZ) effect \cite{Hoscheit:2018nfl}. The temperature perturbation in direction $\hat{n}$ induced by a local void follows
\begin{align}\label{eq:kSZ}
	\Delta T_\mathrm{kSZ}(\hat{n}) = T_\mathrm{CMB} \int_0^{z_e} \delta_e(\hat{n}, z) \frac{V_H(\hat{n}, z) \cdot \hat{n}}{c} \frac{d \tau_e}{dz} dz~.
\end{align}
Here, $T_\mathrm{CMB} = 2.73 \, K$, $\delta_e$ is the density contrast of electrons, and $\tau_e$ is the optical depth along the line of sight. Follow \cite{zhang2015testing}, we set $z_e = 100$ and assume
\begin{align}\label{eq:pecular_vel}
	V_H \simeq [\tilde{H}(t(z), r(z)) - \tilde{H}(t(z), r(z_e))] R(t(z), r(z))~,
\end{align}
where $\tilde{H} = \dot{R}'/R'$. The relation between the optical length and redshift can be described as
\begin{align}\label{eq:optical_depth}
	\frac{d\tau_e}{dz} = \sigma_T n_e(z) c \frac{dt}{dz} = \frac{\sigma_T \theta^2 f_b (1 - Y_\mathrm{He}/2) \Omega_M [1+\delta(r(z))]}{24 \pi G m_p} c \frac{dt}{dz}~.
\end{align}
Here, $\sigma_T$ is the Thomson cross section, $f_b$ is the baryon fraction, $Y_\mathrm{He}$ is the helium mass fraction, $m_p$ is the proton mass, and $\theta$ is defined as $\theta \equiv \tilde{H} + 2H$. Then we can apply Eqs.~\eqref{eq:pecular_vel} and \eqref{eq:optical_depth} in Eq.~\eqref{eq:kSZ}, and use the Limber approximation to obtain the linear kSZ multipole power $C_\ell$ as follows,
\begin{align}
	D_\ell = \frac{\ell (\ell + 1)}{2 \pi}C_\ell \simeq 8 \pi \frac{\ell (\ell + 1)}{(2 \ell + 1)^3} \int_0^{z_e} dz \frac{dr}{dz} r(z) \left(\frac{V_H(r)}{c} \frac{d \tau_e}{dz} \frac{dz}{dr}\right)^2 P_\delta \left(\frac{2 \ell + 1}{2 r(z)}, z\right)~.
\end{align}
Here, $P_\delta(k, z)$ is the $\Lambda$CDM matter power spectrum, which is calculated from the CAMB code \cite{Lewis:1999bs}.

\begin{figure}[ht]
	\centering
	\includegraphics[width=0.46\textwidth]{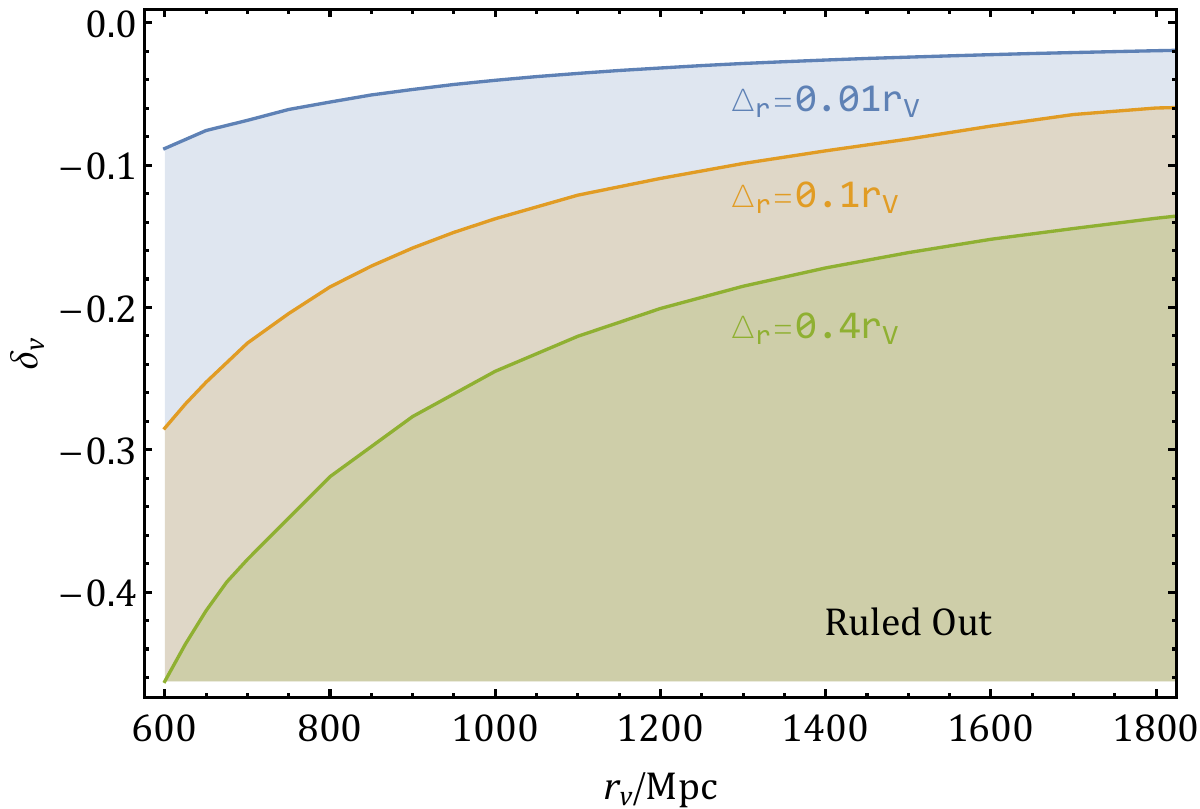}
	\caption{The kSZ constraints on void profiles, see also \cite{Ding:2019mmw}. $\delta_V$ is the matter density contrast, $r_V$ is void radius and $\Delta_r$ is the thickness of the void boundary. The shadow regions are ruled out parameter regions with boundary thicknesses $\Delta_r/r_V = 0.01, 0.1, 0.4$, which is constrained by the requirement $T_\mathrm{CMB}^2 D_\mathrm{3000} < 2.9 \mu K^2$ . We observe that voids with thicker boundaries are significantly less constrained by the kSZ effect.}
	\label{fig:ksz}
\end{figure}

The observed quantity $T_\mathrm{CMB}^2 D_\ell$ at $\ell = 3000$ is bounded by $T_\mathrm{CMB}^2 D_\mathrm{3000} = 2.9 \pm 1.3 \mu K^2$ \cite{George:2014oba} can be used to put strong constraints on void profiles, which is shown in Fig.~\ref{fig:ksz}. 
It shows a Gpc-scale void can hardly have a significant depth under current observational kSZ limit. However, such a constraint could be eased by increasing the width of the void boundary, this is because the bulk flow electrons in a thick edge void have smaller peculiar velocity than those in a sharp edge void. Therefore, collisions between bulk flow electrons and CMB photons produce weaker temperature distortions in the CMB spectrum, which can be found in Eqs.~\eqref{eq:kSZ} and \eqref{eq:pecular_vel} that a smaller $V_H$ can effectively decrease kSZ temperature distortion and further ease the constraints on void depth.

\section{Cosmic Dipoles}

As we have introduced above, the existence of a local void would significantly impact cosmological observations and change the measurement of cosmological parameter, such as Hubble parameters. Cosmic dipoles are another important parameters that would be affected by a local void, due to our off-center position inside the void, the observed spacetime is anisotropic and can induce dipole structure in various cosmic structures, such as CMB temperature dipole at redshift $z = 1100$, $\mathcal{D}_\mathrm{CMB} = (1.23357\pm0.00036) \times 10^{-3}$ \cite{Planck:2018nkj} and quasars number dipole at redshift $z = 1.2$, $\mathcal{D}_\mathrm{quasar} = (1.48\pm0.16) \times 10^{-2}$ \cite{Secrest:2022uvx}. In the framework of FRW cosmology, the CMB dipole and quasar dipole are expected to be explained by the peculiar motion of solar system in CMB frame, however, a $4.9 \sigma$ tension appears in this kinematic interpretation explanation \cite{Secrest:2020has}, which demands us to review the cosmological principle as a basic assumption of FRW cosmology. In the local void scenario,  an off-center position could cause the observations of various anisotropic spacetime backgrounds at different redshifts and produce inconsistent dipole structures in observations, which cause an observational dipolar tension in the framework of FRW cosmology. 

In the following part, we focus on the CMB dipole and quasar dipole induced by the off-center observation inside a Gpc-scale void, and explain their observational dipolar tension in the void cosmology scenario. The void is modeled by the Lemaitre-Tolman-Bondi (LTB) metric as Eq.~\eqref{eq:LTB_metric} and its void profile follows Eq.~\eqref{eq:void_profile} and the parameter of void profile we consider in this work is $r_V = 3400 \, \mathrm{Mpc}$, $\Delta_r = 0.12 r_V$ and $\delta_V = -0.05$, which is allowed under the BAO and kSZ constraints.

\subsection{CMB Dipole}

For an off-center observer, there are two different contributions of the dipole induced by the void. The first originates from the photon trajectories bended by the local void structure, and the second comes from the "peculiar velocity" created by the different values of the Hubble parameter in different locations of the void model. These two dipole contributions are of the same order, therefore we should consider them both.  We will now discuss in detail how the whole dipole is calculated.

\begin{figure}[ht]
    \centering
	\includegraphics[width=0.46\textwidth]{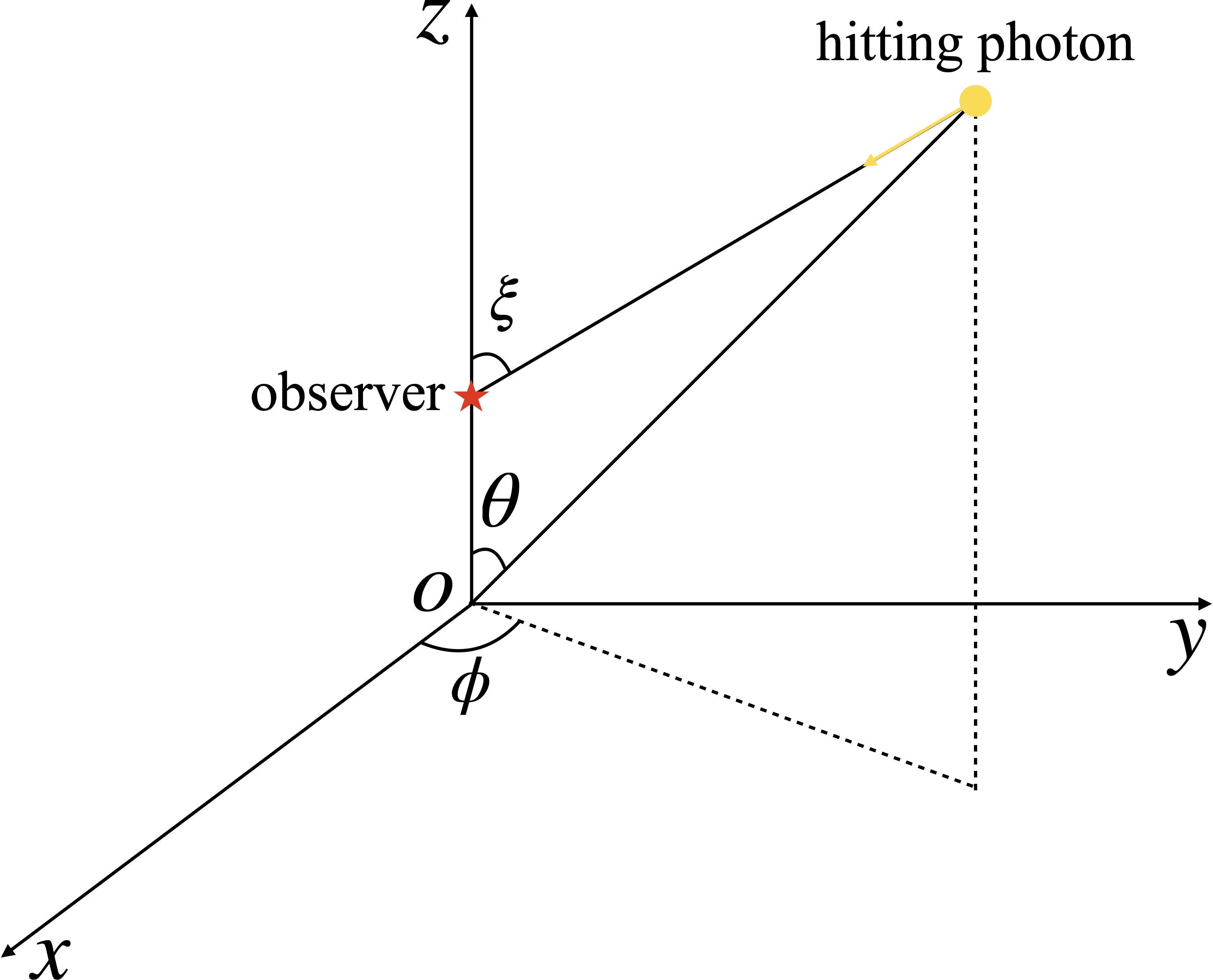}
	\caption{This figure demonstrates the coordinate system in calculation. The origin of coordinate $O$ is the center of the local void, the red star denotes the observer with distance $d$ to void center, and yellow dot denotes the hitting photon to the observer with an observational angle $\xi$.}
	\label{coordinate}
\end{figure}

For the first part of the dipole, it is crucial to specify the photon trajectories. We follow \cite{Alnes:2006pf} to start from the geodesic equation of photons. Due to our off-center position inside the void, the photon trajectories should follow axial symmetry and be independent of the azimuth angle $\phi$, then we have the following  geodesic equations for spacetime coordinates $(t, r, \theta)$,
\begin{align}
	\frac{d^2 t}{d \lambda^2} + \frac{R' \dot{R}'}{1 - k} \left(\frac{dr}{d\lambda}\right)^2 + R \dot{R} \left(\frac{d\theta}{d\lambda}\right)^2 = 0~,\\
	\frac{d^2 r}{d \lambda^2} + \left(\frac{R''}{R'} + \frac{k'}{2 - 2 k}\right) \left(\frac{dr}{d\lambda}\right)^2 + 2 \frac{\dot{R}'}{R'} \frac{dr}{d \lambda} \frac{dt}{d \lambda} - \frac{R (1 - k)}{R'} \left(\frac{d\theta}{d\lambda}\right)^2 = 0~,\\
	\frac{d^2 \theta}{d \lambda^2} + 2 \frac{R'}{R} \frac{d \theta}{d \lambda} \frac{dr}{d \lambda} + 2 \frac{\dot{R}}{R} \frac{d \theta}{d \lambda}\frac{dt}{d \lambda} = 0~.
\end{align}
Here, $\lambda$ is a parameter defined along the trajectories of photons and $k$ is spacetime curvature.
Given the position of the observer and the direction of photon arrival as initial conditions, we can numerically solve the geodesic equations of photons backward in time from the present cosmic time to a past cosmic time. After obtaining the solution of the geodesic equation of photons, the cosmic redshift along the trajectory of photon $z(\lambda)$ can be calculated by solving the following equation, 
\begin{align}
	\frac{dz}{d\lambda} = -(1+z) \frac{d \lambda}{dt} \left[\frac{R' \dot{R}'}{1 - k} \left(\frac{dr}{d\lambda}\right)^2 + R \dot{R}\left(\frac{d\theta}{d\lambda}\right)^2\right]~.
\end{align}
Then we can calculate the redshift distribution of photons on observational direction $\hat{n}$ from a constant time hypersurface $z(\hat{n}, t)$ \cite{Alnes:2006pf}. Assuming that the observer is placed in the $z$-axis with distance $d$. For a photon hitting the observer at angle $\xi$ relative to the $z$-axis, as demonstrated in Fig~\ref{coordinate}, the corresponding CMB temperature is then directly related to redshift
\begin{equation}\label{eq:temp_red}
	T(\xi)=\frac{T_*}{1+z(\xi)}~,
\end{equation}
where $T_*$ is the temperature at the last-scattering surface. 
Then the relative temperature variation is 
\begin{equation}\label{eq:vari}
	\Theta(\xi)\equiv\frac{\Delta T}{\widehat{T}}=\frac{T(\xi)-\widehat{T}}{\widehat{T}}~,
\end{equation}
with the average temperature $\widehat{T}=\int\mathrm{d}\Omega\,T(\xi)/4\pi$. Now we can calculate the amplitude of this dipole contribution,
\begin{equation}\label{eq:dipole}
	\mathcal{D}_1=\frac{2}{\pi}\int^{\pi}_0\Theta(\xi)\cos\xi\,\mathrm{d}\xi~.
\end{equation}

There is one more contribution from the void to the final dipole. As we are located in an off-center position, we will experience a pull from the center of the void due to its structure. This pull results in a ``peculiar velocity'' that causes our off-center observer to move away from the center of the void. This velocity can be calculated similarly as in Eq.~\eqref{eq:pecular_vel}, and the second part of the dipole, as defined as the void pull term, is then 
\begin{align}\label{eq:extra_pull}
	\mathcal{D}_2= \frac{v_H(d,z)}{c} \simeq\frac{1}{c} [\tilde{H}(t_0, d) - \tilde{H}(t_0, r(z))] R(t_0, d)~,
\end{align}
where $\tilde{H}=\dot{R}'/R'$. $t_0=t_B$ is the cosmic time since the big bang, $d$ is the distance of the observer to the void center, $z$ is the redshift corresponding to the dipole we are observing. Since the directions of these two dipole contributions are identical, that is the direction of the off-center observer relative to the center of the void, we can obtain the amplitude of the overall dipole induced by the void model by simply adding the contributions up
\begin{equation}\label{eq:fulldipole}
    \mathcal{D}=\mathcal{D}_1+\mathcal{D}_2=\frac{2}{\pi}\int^{\pi}_0\Theta(\xi)\cos\xi\,\mathrm{d}\xi+\frac{v_H(d,z)}{c}~.
\end{equation}

\begin{figure}[ht]
    \centering
	\includegraphics[width=0.66\textwidth]{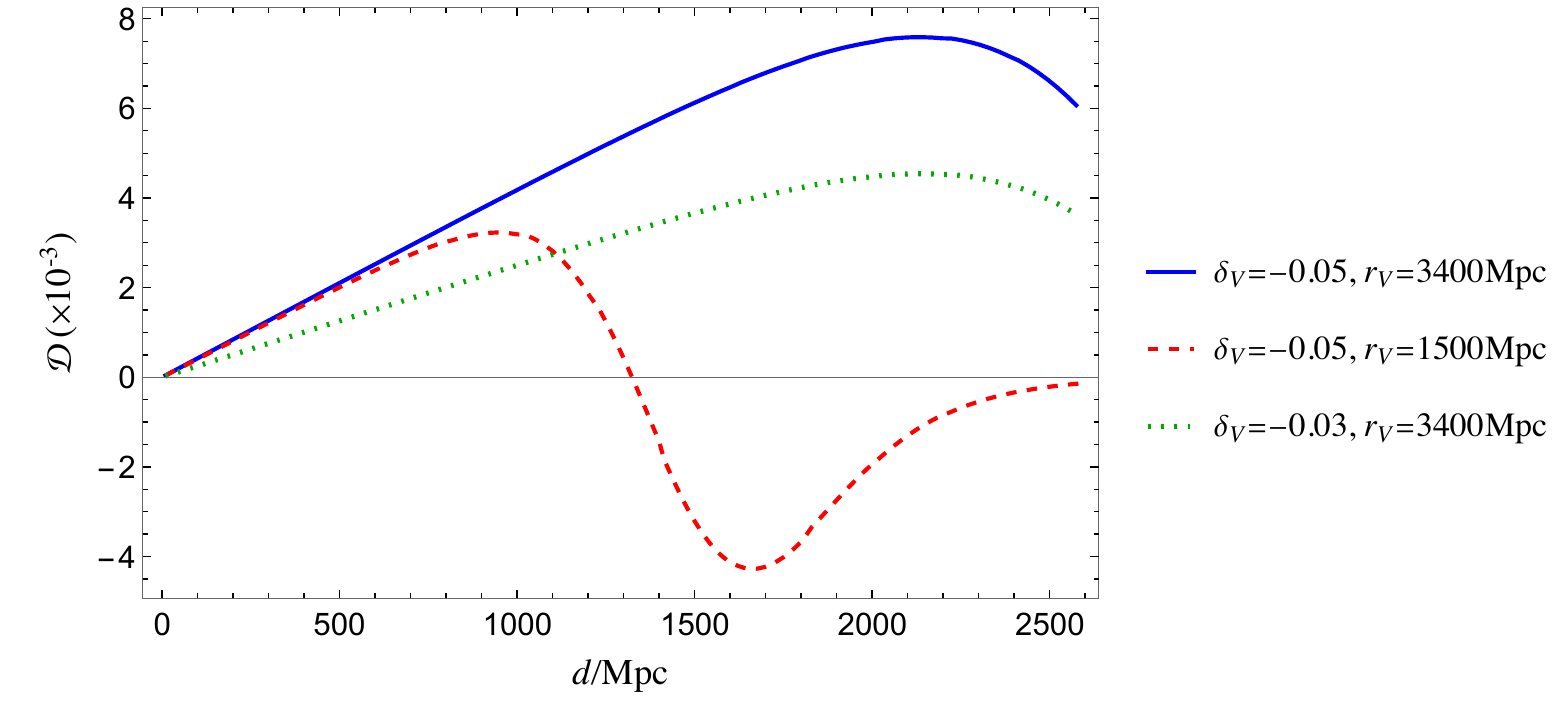}
	\caption{CMB dipole as a function of observer's location, where $d$ is the distance from the observer to the void center. Different sets of parameters of the void are shown in the labels with the same relative edge width $\Delta_r=0.12r_V$.}
	\label{ad}
\end{figure}

\begin{figure}[htbp]
	\centering
	\includegraphics[width=0.46\textwidth]{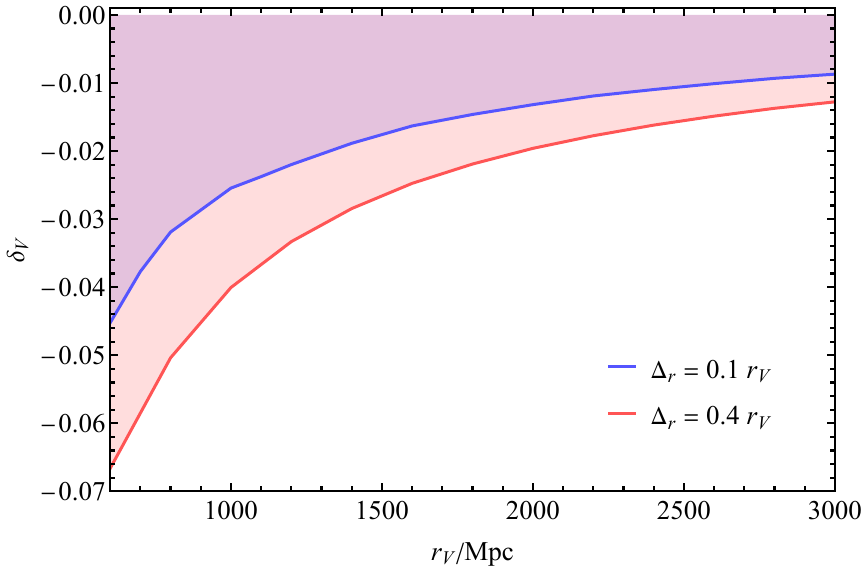}
        \includegraphics[width=0.45\textwidth]{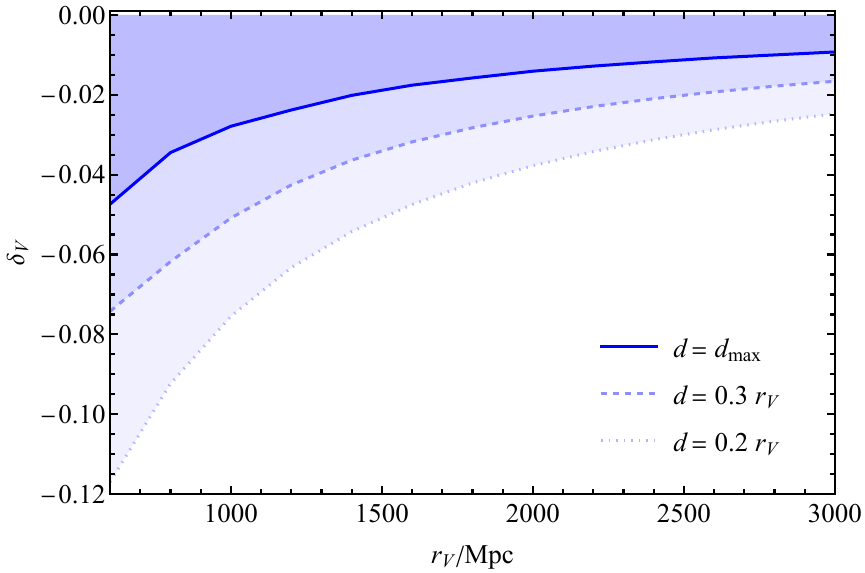}
	\caption{\textit{Left panel:} Void profile constraints from CMB dipole for different void thickness, where constrained shadow regions statisfy $\mathcal{D}_\mathrm{max} < \mathcal{D}_\mathrm{CMB}$, and different colors denote the constraints with various void thickness. \textit{Right panel:} Void profile constraints from CMB dipole for different locations of the observer, where constrained shadow regions statisfy $\mathcal{D}(d = d_\mathrm{label}) < \mathcal{D}_\mathrm{CMB}$. The types of curve denote the different locations of the observer and $d_\mathrm{max}$ denotes the location of the observer that can produce the maximal dipole at $z_\mathrm{CMB}$. The void thickness is set to $\Delta_r = 0.12 r_V$.}
	\label{fig:dipole_constraint}
\end{figure}

Naturally, the distance from the off-center observer to the center of the void plays an important role. As shown in Fig.~\ref{ad}, the CMB dipole first increases as the observer location becomes further away from the void center, but when the observer gets closer to the void boundary, the dipole reaches its maximum and starts to decrease, even to a negative value. The change in dipole direction around the void boundary comes from the void pull term $\mathcal{D}_2=v_H/c$, which starts to become negative around the void boundary. As the observer passes the void boundary and gets even further, the absolute value of the dipole decreases and converges to zero, which is conceivable since the observer is no longer inside the void. Meanwhile, the void profile is also significant to the observed CMB dipole. Comparing different curves in Fig.~\ref{ad}, when the observer is inside the dipole, the amplitude of the observed CMB dipole is generally smaller for a shallower void, and a larger void radius could produce a larger anisotropic region in the universe. We can give a brief constraint to rule out some void profiles, if the maximal value of the induced dipole at $z_\mathrm{CMB}$ is smaller than the observed CMB dipole $\mathcal{D}_\mathrm{max} < \mathcal{D}_\mathrm{CMB}$, 
which is shown in the left panel of Fig.~\ref{fig:dipole_constraint}. Also, we can constrain the void profile based on the location of the observer, that is, for a given location of the observer inside the void, which kinds of void profile cannot produce the observed CMB dipole. This requirement can be expressed as $\mathcal{D}(d) < \mathcal{D}_\mathrm{CMB}$, the constrained regions are shown in the right panel of Fig.~\ref{fig:dipole_constraint}.

In the left panel of Fig.~\ref{fig:dipole_constraint}, we can find that, to observe a CMB dipole, a smaller size void requires a deeper void depth; meanwhile, a larger size void only needs a shallower depth. This is consistent with our understanding, for two void profiles with the same depth and thickness, but different void radius, a larger void radius can produce a larger CMB dipole, shown by the blue solid curve and red dashed curve in Fig.~\ref{ad}. In the right panel of Fig.~\ref{fig:dipole_constraint}, we can find that, in order to obtain the observed value of the CMB dipole, if we are close to the void center, the void depth must be large to achieve the CMB dipole value. It also shows that our distance to the void center strongly depends on the void profile.

As in the previous setting, we consider the parameter of void profile $(r_V, \Delta_r, \delta_V) = (3.4 \, \mathrm{Gpc}, 0.41 \, \mathrm{Gpc}, -0.05)$. In this benchmark model, the amplitude of detected CMB dipole $\mathcal{D} \simeq 1.23 \times 10^{-3}$ \cite{Planck:2018nkj} corresponds with the location of the observer, which is $292 \, \mathrm{Mpc}$ away from the center of the void, which we will use in the following discussion. We need to pay attention to the location of the observer, it is a $8.6\%$ offset from the void center. Although it is relatively close to the void center, it does not have the coincidence problem, as we have emphasized in the right panel of Fig.~\ref{fig:dipole_constraint}. The distance to the void center depends on void profile, and this $8.6\%$ offset is only the value in our benchmark model; a realistic location of the observer needs future observations on this void profile.

\begin{figure}[htbp]
	\centering
	\includegraphics[width=0.46\textwidth]{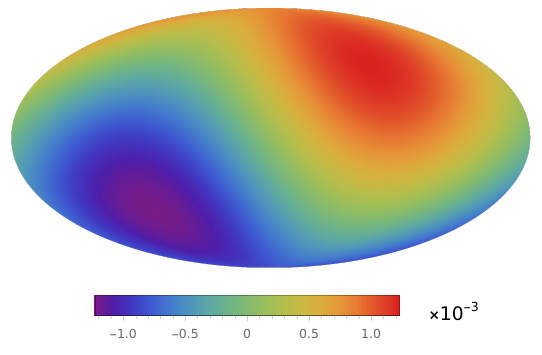}
	\caption{Observed CMB dipole in the view of an off-center observer. The distance of the observer from void center is set to be $292 \, \text{Mpc}$ and the position direction from void center to the observer is set to (264\textdegree, 48\textdegree) in galactic coordinates, which is coincide with measured CMB dipole direction.}
	\label{dip}
\end{figure}

As the LTB metric is isotropic, the orientation of the CMB dipole seen by an off-center observer is the same as the direction from the void center to the observer. We can simply set the position direction of the off-center observer to be consistent with the CMB dipole, which is $(l,b) = $ (264\textdegree, 48\textdegree) in galactic coordinate \cite{Planck:2018nkj}. Fig.~\ref{dip} shows an example plot of the CMB dipole seen by an observer located $292 \, \mathrm{Mpc}$ from the center.

\begin{figure}[htbp]
	\centering
	\includegraphics[width=0.46\textwidth]{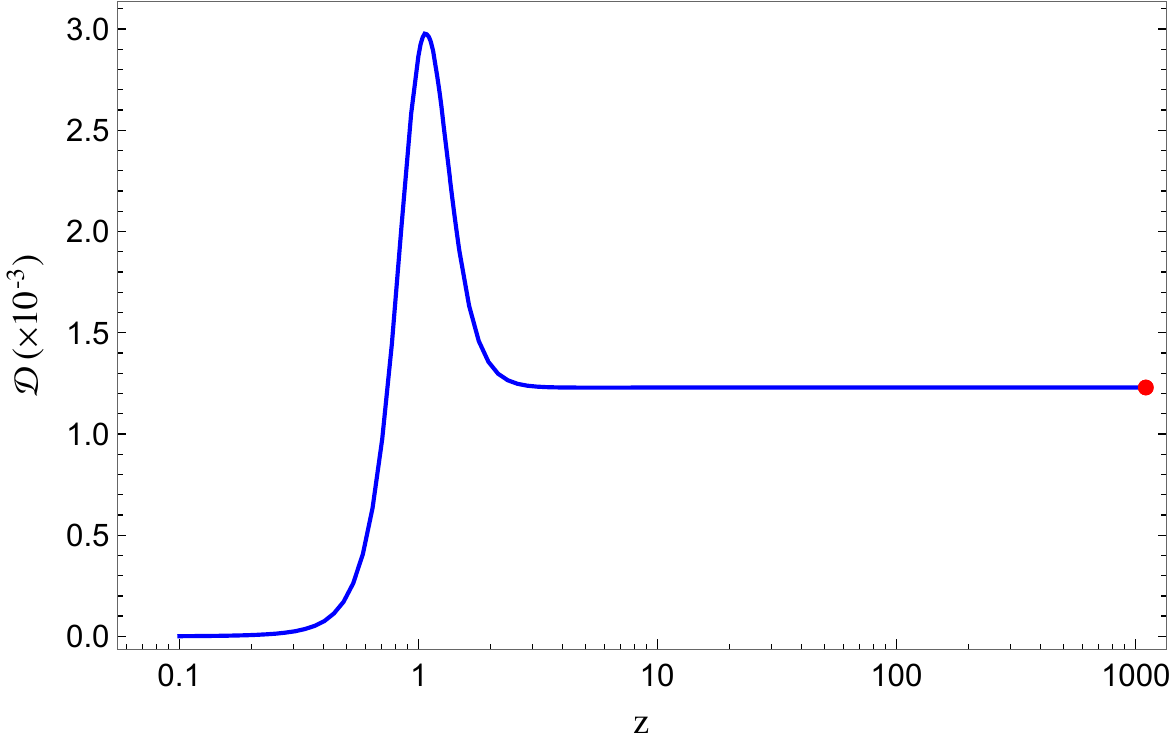}
	\caption{Overall induced redshift dipole in the view of an off-center observer. The distance of the observer from void center is set to be $292 \, \text{Mpc}$ and void profile is $(r_V, \Delta_r, \delta_V) = (3.4 \, \mathrm{Gpc}, 0.41 \, \mathrm{Gpc}, -0.05)$. The red data point is observed CMB dipole $\mathcal{D} = (1.23357 \pm 0.00036) \times 10^{-3}$ at redshift $z = 1100$ \cite{Planck:2018nkj}.}
	\label{fig:cmbdip_z}
\end{figure}

Note that this method can be extended from CMB where $z=1100$ to cosmic dipoles corresponding to lower redshifts, as long as the observable signal is related to redshift in the same way as temperature in Eq.\eqref{eq:temp_red}; such dipole relation is shown in Fig.~\ref{fig:cmbdip_z}.
We can find that the amplitude of induced cosmic dipole reaches maximum around $z=1$, just at the edge of the void. This is because the matter density contrast achieves its maximal value around the void boundary, which strongly increases the spacetime anisotropy and enhances the dipole structure in the redshift distribution of observable signals. When the source gets away from the void, the amplitude of the dipole becomes stable, which corresponds with measured CMB dipole $\mathcal{D} \simeq 1.23 \times 10^{-3}$, since the source can be approximately considered to be at infinity.

\subsection{Quasar Dipole}

Besides CMB, distant sources like quasars can also form dipoles in void cosmology. For CMB, the dipole appears in its temperature anisotropy, while for quasar, the dipole appears in number counting anisotropy. That is to say angular number density $\text{d}N/\text{d}\Omega$ is different when the observer looks into different directions $\xi$.

Such anisotropy can be derived using similar methods in the kinematic interpretation of the quasar dipole \cite{1984MNRAS.206..377E}. Consider an off-center observer in a local void; the frequency of observed photons $\nu_\text{o}$ is redshifted from the emission frequency $\nu_\text{e}$ as
\begin{equation}\label{eq:nu_n}
	\nu_{\text{o}}(\xi)=\frac{\nu_{\text e}}{1 + z(\xi)}~.
\end{equation}
As Eq.~\eqref{eq:vari}, we can calculate the variance of observed frequency of photons as,
\begin{equation}
	\Theta \equiv \frac{\Delta \nu_\text{o}}{\widehat{\nu}_\text{o}} = \frac{\nu_\text{o}(\xi) - \widehat{\nu}_\text{o}}{\widehat{\nu}_\text{o}}~,
\end{equation}
where $\widehat{\nu}_\text{o} = \int d \Omega \, \nu_\text{o}(\xi)/4 \pi$, then we define mean redshift $\bar{z} \equiv \nu_\text{e}/\widehat{\nu}_\text{o}$, which transforms Eq.~\eqref{eq:nu_n} into
\begin{equation}
	\nu_\text{o} = \widehat{\nu}_\text{o} \kappa(\xi)~,
\end{equation}
where $\kappa(\xi)=(1+\bar{z})/(1+z(\xi))$. Such an angular-dependent frequency shift is induced by the redshift anisotropy in an off-center void. This causes the anisotropy in power-law spectral energy distribution of the source $S \propto \nu^{-\alpha}$ and a cumulative power-law distribution above a limiting apparent flux density $N(>S) \propto S^{-x}$. The observed solid angle can be approximated to $d\Omega_o \simeq d\Omega_e \kappa(\xi)^{-2}$ \cite{Hogg:1999ad}. Accordingly, the observed angular number density is
\begin{equation}\label{eq8}
	\Big(\frac{\text{d}N}{\text{d}\Omega}\Big)_{\text o}\simeq\Big(\frac{\text{d}N}{\text{d}\Omega}\Big)_{\text e}\kappa(\xi)^\gamma~,
\end{equation}
where the index $\gamma = 2 + x(1 + \alpha)$. Following \cite{Secrest:2022uvx}, we set $\alpha = 1.06$, $x = 1.89$. The averaged angular number density can be integrated
\begin{equation}\label{eq9}
	\Big(\frac{\text{d}N}{\text{d}\Omega}\Big)_{\text a}=\frac{1}{4\pi}\int\Big(\frac{\text{d}N}{\text{d}\Omega}\Big)_{\text o}\text d\Omega~.
\end{equation}
The relative variation is
\begin{equation}
	\Theta(\xi)=\frac{\Delta(\text d N/\text d\Omega)}{(\text d N/\text d\Omega)_a}=\frac{(\text d N/\text d\Omega)_{\text o}-(\text d N/\text d\Omega)_a}{(\text d N/\text d\Omega)_a}~,
\end{equation}
then the amplitude of the corresponding dipole $\mathcal{D}_1$ can be calculated as Eq.~\eqref{eq:dipole}. For the void pull term $\mathcal{D}_2$, the quasar dipole contribution is slightly different from the CMB case \cite{1984MNRAS.206..377E}
\begin{equation}\label{eq:extra pull quasar}
    \mathcal{D}_2=\gamma\frac{v_H(d,z)}{c}=\left[2+x(1+\alpha)\right]\frac{v_H(d,z)}{c}.
\end{equation}
The above calculation gives a quasar number dipole $\mathcal{D}=\mathcal{D}_1+\mathcal{D}_2=1.63\times10^{-2}$ at redshift $z\simeq1.2$ for an observation at the location of 292 Mpc away from the void center with void profile $(r_V, \Delta_r, \delta_V) =  (3.4\, \mathrm{Gpc}, 0.41 \, \mathrm{Gpc}, -0.05)$. This dipole value is within $1 \sigma$ range of the quasar dipole $\mathcal{D} = (1.48 \pm 0.16) \times 10^{-2}$ from the CatWISE data \cite{Secrest:2022uvx}.

Similar to the temperature dipole, this dipole comes from $\kappa(\xi)$ term, due to redshift distribution in different directions. Just as redshift dipole behavior in Fig.~\ref{fig:cmbdip_z}, it is clear that there is a peak for the dipole anisotropy around $z = 1$, due to the existence of a Gpc-scale local void. Therefore, when considering purely kinematic interpretations of quasar dipoles, such a peak dipole anisotropy induced by an off-center void can cause a larger amplitude in the quasar dipole. This reconciles the peculiar velocity inconsistency between the quasar and CMB dipoles.

\begin{figure}[htbp]
    \centering
        \includegraphics[width=0.46\textwidth]{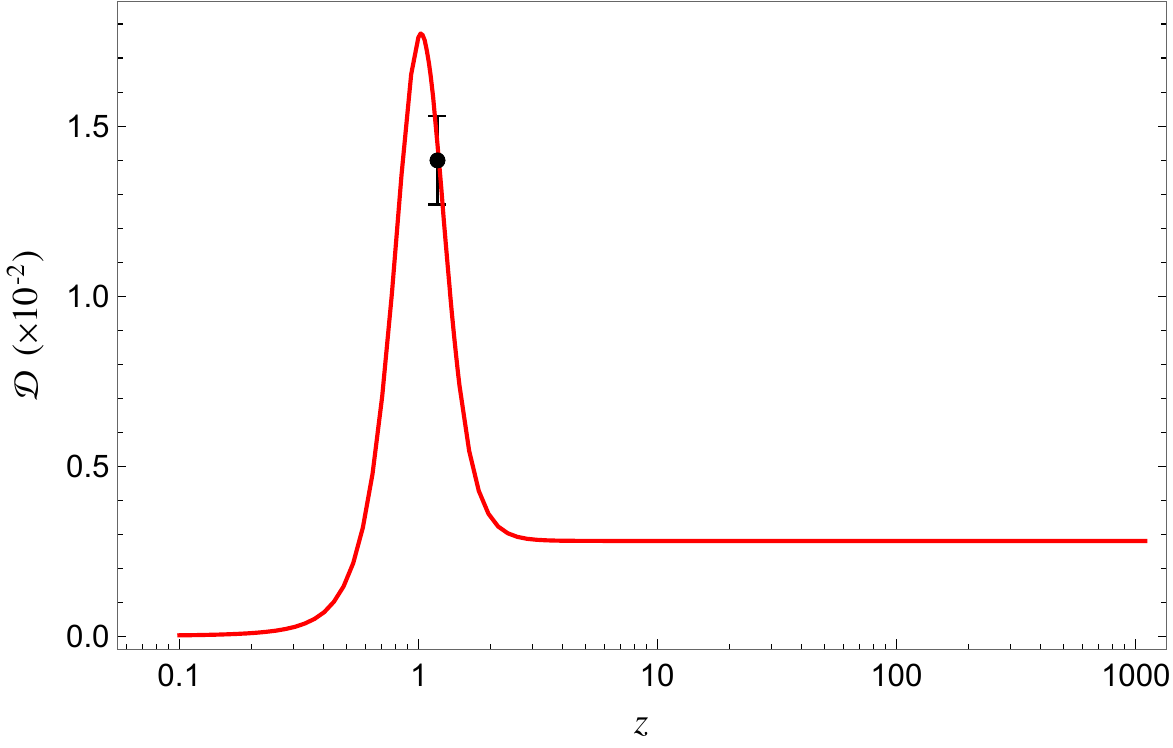}
	\caption{Overall angular matter distribution dipole as a function of redshift. The observer is located at $292 \, \text{Mpc}$ from the void center and void profile is $(r_V, \Delta_r, \delta_V) = (3.4 \, \mathrm{Gpc}, 0.41 \, \mathrm{Gpc}, -0.05)$. The black data point is the quasar dipole $\mathcal{D} = (1.48 \pm 0.16) \times 10^{-2}$ at the mean redshift of quasars $z = 1.2$ \cite{Secrest:2022uvx}.}
	\label{cmbaz}
\end{figure}

In principle, there is an intrinsic matter distribution anisotropy in the view of an off-center observer inside the void, and it is the origin of the observed quasar dipole. If we assume that the number density of quasars is determined by their surrounding matter density, the amplitude of the quasar dipole should be similar to the amplitude of the angular matter density dipole. In calculating angular matter density in a local void, the mass term is the matter density times the observed volume elements and the angular term is the observed solid angle, which can be expressed as follows, 
\begin{equation}\label{eq11}
	\frac{\text d M}{\text d\Omega}(\xi) = \frac{\rho\text dS\text dt}{\text d\Omega}=\frac{\rho\sqrt{\frac{R'^2}{1-k}\text dr^2+R^2\text d\theta^2}R\sin{\theta}\text d\phi\text dt}{\sin{\xi}\text d\xi\text d\phi}=\rho\sqrt{\frac{R'^2}{1-k}\Big(\frac{\text dr}{\text d\xi}\Big)^2+R^2\Big(\frac{\text d\theta}{\text d\xi}\Big)^2}R\frac{\sin{\theta}}{\sin{\xi}}\text dt,
\end{equation}
where $\text dS=\sqrt{R'^2/(1-k)\text dr^2+R^2\text d\theta^2}R\sin{\theta}\text d\phi$ is the area element perpendicular to the trajectory direction of the hitting photon, and $\text dt$ is the physical time interval between two nearby constant time hypersurfaces. Relative variation $\Theta$ and the corresponding dipole amplitude $\mathcal{D}_1$ can be therefore calculated following Eq.~\eqref{eq:dipole}. The void pull term is still $\mathcal{D}_2=\gamma v_H/c$ for quasar as previously shown. The results of the overall angular matter distribution dipole $\mathcal{D}=\mathcal{D}_1+\mathcal{D}_2$ are shown in Fig.~\ref{cmbaz}.

In Fig.~\ref{cmbaz}, we can find the angular matter density produces a dipole value $\mathcal{D} \simeq 1.48 \times 10^{-2}$ in the local void scenario, which is consistent with the observed quasar number dipole within $1 \, \sigma$ at redshift $z = 1.2$, which is the mean redshift of observed quasars \cite{Marocco_2021}. Also, compare with the above calculated quasar dipole based on redshift anisotropy dipole, the amplitude of these two dipolar anisotropies takes similar values around $z = 1.2$; it shows the angular matter density dipole could be a decent approximation to the quasar number dipole. However, due to ignoring the curved trajectories of photons, the approach to calculate quasar dipole via redshift anisotropy distribution cannot extend to the region near and outside the local void, where Eq.~\eqref{eq11} needs to be considered.

\begin{figure}[h]
    \centering
        \includegraphics[width=0.46\textwidth]{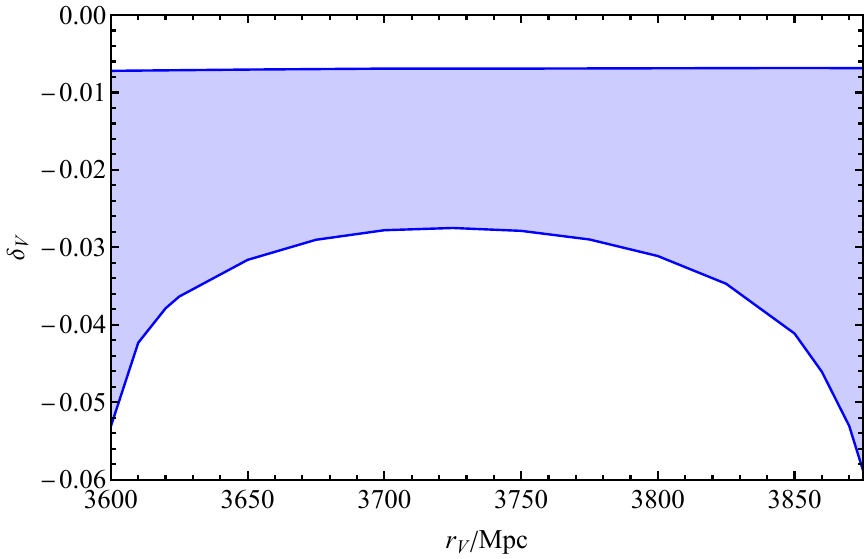}
	\caption{This figure demonstrates the allowed parameter regions of void profile that can match the observed CMB temperature dipole and quasar number dipole. The void thickness is set as $\Delta_r = 0.12 r_V$.}
	\label{allow_region}
\end{figure}

Based on the above discussion, we can find the allowed parameter ranges of the local void profile, whose induced dipole should satisfy the temperature dipole at the redshift of CMB should be around $1.23357 \times 10^{-3}$ and angular matter density dipole at the mean redshift of observed quasars should be in the range of $(1.48 \pm 0.16) \times 10^{-2}$ as follows,
\begin{align}
    \mathcal{D}_T(z_\mathrm{CMB}) = (1.23357\pm0.00036) \times 10^{-3}~,~~~\mathcal{D}_A(z_\mathrm{quasar}) = (1.48 \pm 0.16) \times 10^{-2}~.
\end{align}
According to this condition, the allowed parameter regions of void profiles are shown in Fig.~\ref{allow_region}, where we set the void thickness as $\Delta_r/r_V = 0.12$. We can find that the allowed range of void depth is shallowest at $r_V \simeq 3720 \, \mathrm{Mpc}$, which corresponds with redshift $z \simeq 1.2$. As we have discussed above, the maximal value of induced dipole appears around the boundary of the local void. If this void boundary corresponds with the mean redshift of observed quasars, the required void depth that matches observed dipoles should be smallest. For a smaller or larger local void, the maximal induced dipole does not appear at the redshift of observed quasars, so they need a larger void depth to explain observed dipoles.

\section{Void considering peculiar velocity}

\subsection{Dipole requirements}

Previously we focus on discussing the ability of the void model in explaining the cosmic dipoles all by itself. In those discussions, even when the observer has a "peculiar velocity", this velocity is induced by the void itself and will vanish once we switch to the standard $\Lambda$CDM model. However observations on relatively local structures have shown that our Local Group has a peculiar velocity which can explain the CMB dipole to a good extent, though not fully. Therefore, it is also meaningful to discuss whether our void model can ease or even fill the gap between the known peculiar velocity and the comic dipoles. The idea of implementing peculiar velocities in our void model is simply to combine dipoles together. Dipole amplitudes induced by void or peculiar velocities are all small, so we can treat the dipole combination as effectively vector addition. Setting the CMB and quasar dipole contributions from the void model as $\boldsymbol{\mathcal{D}}_\text{c},\boldsymbol{\mathcal{D}}_\text{q}$, we then expect they each can satisfy the observed dipoles after being combined with contributions from the reconstructed peculiar velocity of the sun in the CMB frame $\mathbf{v}_s$
\begin{align}
    \boldsymbol{\mathcal{D}}_{\text{c}}+\frac{\mathbf{v}_{\text s}}{c}=&\boldsymbol{\mathcal{D}}_{\text{CMB}},\\
    \boldsymbol{\mathcal{D}}_{\text{q}}+\gamma\frac{\mathbf{v}_{\text s}}{c}=&\boldsymbol{\mathcal{D}}_{\text{quasar}}.
\end{align}
where for CMB dipole $|\boldsymbol{\mathcal{D}}_\text{CMB}|=(1.23357\pm0.00036)\times10^{-3},(l,b)=(264\dotdeg021\pm0\dotdeg011,48\dotdeg253\pm0\dotdeg005)$ \cite{Planck:2018nkj}, quasar dipole $|\boldsymbol{\mathcal{D}}_\text{quasar}|=(1.48\pm0.16)\times10^{-2},(l,b)=(238^\circ\pm7^\circ,31^\circ\pm5^\circ)$ \cite{Secrest:2022uvx}.

Here, $\mathbf{v}_s$ is not the peculiar velocity directly inferred from CMB; if it were, $\boldsymbol{\mathcal{D}}_c$ would be just 0. $\mathbf{v}_s$ is obtained from data sets of relatively local structures $\mathbf{v}_s=\mathbf{v}_{s-LG}+\mathbf{v}_{LG}$, where $\mathbf{v}_{s-LG}$ is the velocity of the sun relative to the Local Group \cite{Planck:2018nkj,Diaz:2014kqa} with $|\mathbf{v}_{s-LG}|=299\pm15\text{km}\cdot\text{s}^{-1},(l,b)=(98\dotdeg4\pm3\dotdeg6,-5\dotdeg9\pm3\dotdeg0)$; $\mathbf{v}_{LG}$ is the peculiar velocity of the Local Group in the CMB frame reconstructed using Cosmicflows-4 data \cite{Tully:2022rbj, Courtois:2022mxo, Stiskalek:2025xaw} with $|\mathbf{v}_{LG}|=766\pm336\text{km}\cdot\text{s}^{-1},(l,b)=(292\dotdeg7\pm35\dotdeg5,11\dotdeg7\pm31\dotdeg2)$. Now we have $|\mathbf{v}_s|=278\pm352\text{km}\cdot\text{s}^{-1},(l,b)=(309^\circ\pm79^\circ,22^\circ\pm80^\circ)$ and become able to calculate the corresponding required cosmic dipoles to fill the gap, CMB $|\boldsymbol{\mathcal{D}}_\text{c}|=(1.07\pm0.92)\times10^{-3},(l,b)=(257^\circ\pm58^\circ,47^\circ\pm48^\circ)$ and quasar $|\boldsymbol{\mathcal{D}}_\text{q}|=(1.41\pm0.51)\times10^{-2},(l,b)=(235^\circ\pm22^\circ,29^\circ\pm24^\circ)$. The numerical simulation method used to calculate the addition of vector distributions is introduced in Appendix \ref{app}. Note that these requirements are independent of the void model. The uncertainty of these dipole requirements is quite large, most comes from the uncertainty in the reconstructed peculiar velocities $\mathbf{v}_{LG}$. When the reconstruction method changes, $\mathbf{v}_{LG}$ is different but still has large uncertainty \cite{Stiskalek:2025xaw}, leading to the fact that $\boldsymbol{\mathcal{D}}_c,\boldsymbol{\mathcal{D}}_q$ are shifted but still with large uncertainty. In this work, we will use the $\mathbf{v}_{LG}$ and $\boldsymbol{\mathcal{D}}_c,\boldsymbol{\mathcal{D}}_q$ chosen and calculated as before.

\subsection{CMB Dipole}

The above analysis of dipole requirement means that our previous calculation of dipole induced by a void model does not need to change since it is the requirement that is changed. We then consider a void profile $(r_V, \Delta_r, \delta_V) = (3.5 \, \mathrm{Gpc}, 0.42 \, \mathrm{Gpc}, -0.058)$. Taking into account the direction constraints of $\boldsymbol{\mathcal{D}}_c$ and $\boldsymbol{\mathcal{D}}_q$, we set the void direction $(l,b)=(240^\circ,33^\circ)$. Similarly, we can calculate the dipoles at different redshifts seen by an off-center observer located 219\,Mpc from the center and see how well it meets the CMB dipole requirement $\boldsymbol{\mathcal{D}}_c$.

\begin{figure}[ht]
	\includegraphics[width=0.46\textwidth]{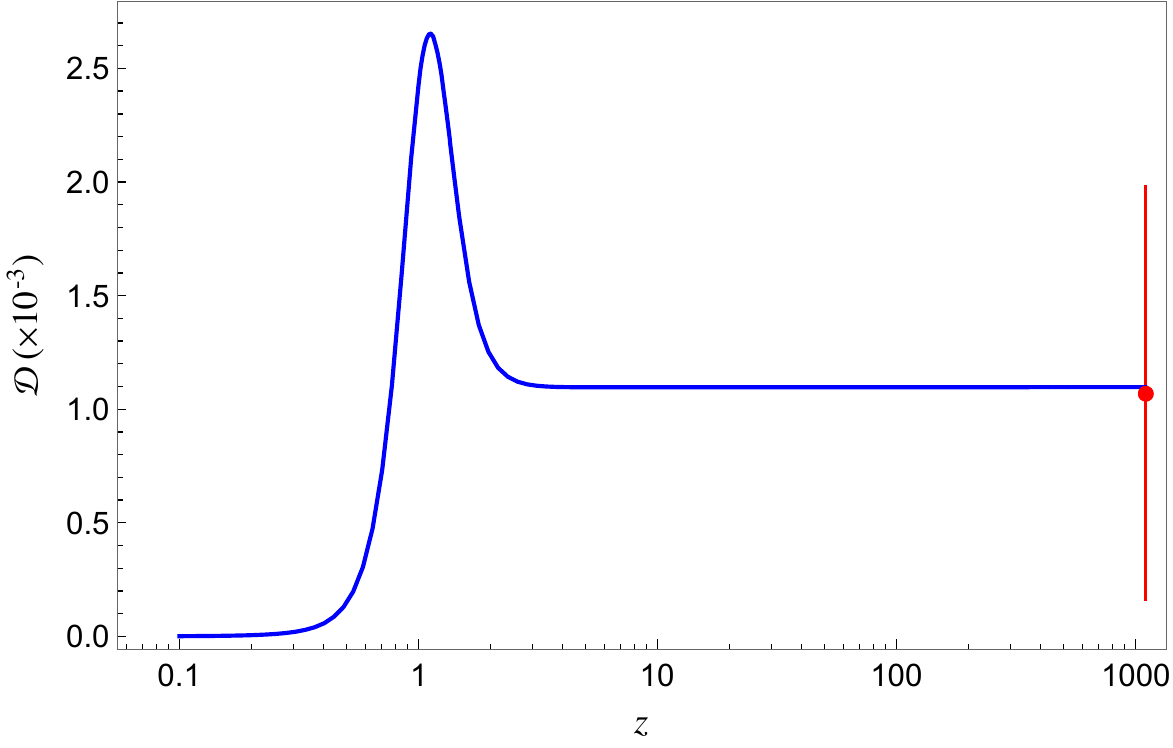}
	\caption{Overall induced redshift dipole in the view of an off-center observer. The distance of the observer from void center is set to be $219 \, \text{Mpc}$ and void profile is $(r_V, \Delta_r, \delta_V) = (3.5 \, \mathrm{Gpc}, 0.42 \, \mathrm{Gpc}, -0.058)$. The red data point and error bar is the CMB dipole requirement when taking into account our observed peculiar velocity $|\boldsymbol{\mathcal{D}}_c| = (1.07 \pm 0.92) \times 10^{-3}$ at redshift $z = 1100$.}
	\label{p2_dpcmb_z}
\end{figure}

Fig.~\ref{p2_dpcmb_z} shows that when we consider the observed peculiar velocity of the solar system,  the dipole induced by our void model can still fit the requirement of the CMB dipole well. Similarly to Fig.~\ref{fig:cmbdip_z}, the dipole reaches a maximum around $z=1$ and becomes stable for larger redshifts. The reason for this similarity is that the calculation methods in these two plots are the same, only the void profile parameters are slightly changed to meet a different requirement.

\subsection{Quasar Dipole}

\begin{figure}[ht]
    \centering
	\includegraphics[width=0.46\textwidth]{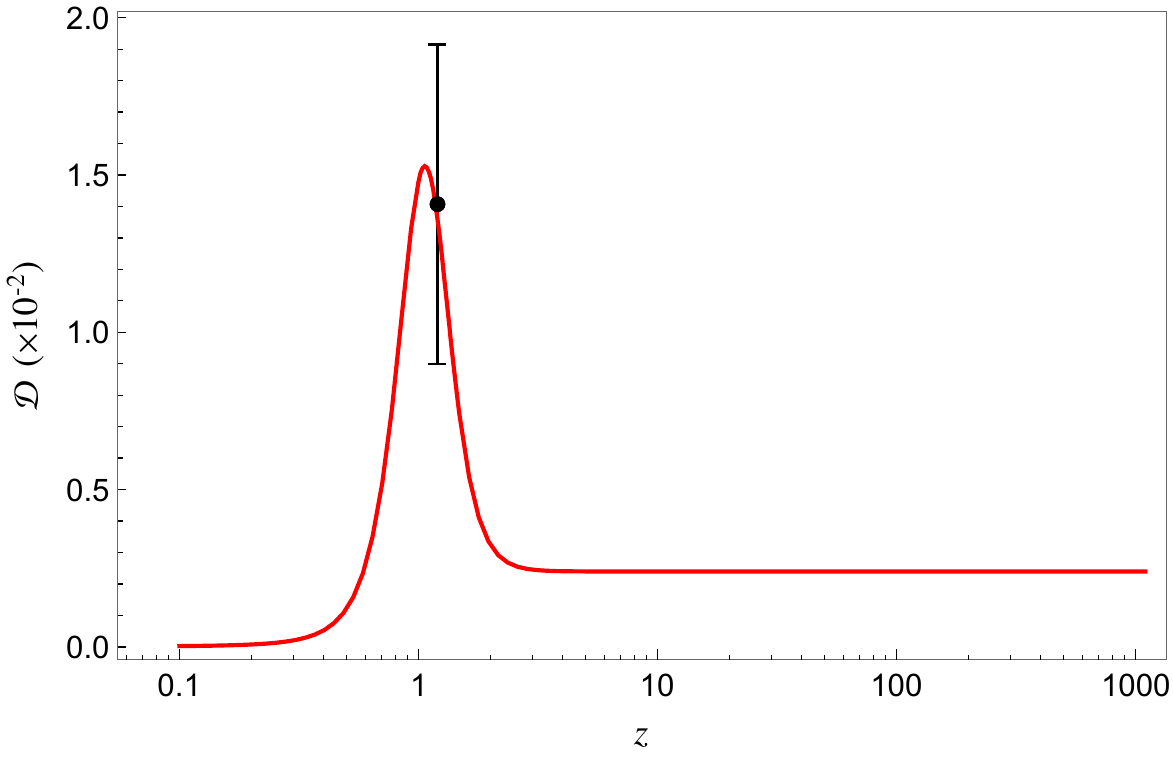}
	\caption{Overall angular matter distribution dipole as a function of redshift. The observer is located at $219 \, \text{Mpc}$ from the void center and void profile is $(r_V, \Delta_r, \delta_V) = (3.5 \, \mathrm{Gpc}, 0.42 \, \mathrm{Gpc}, -0.058)$. The black data point is the quasar dipole requirement for the void $|\boldsymbol{\mathcal{D}}_q| = (1.41 \pm 0.151) \times 10^{-2}$ at the mean redshift of quasars $z = 1.2$.}
	\label{p2_dpqua_z}
\end{figure}

We follow the same method in previous sections, calculating the dipole induced by angular matter density $\text d M/\text d\Omega$ and combining it to the void pull term $\mathcal{D}_2=\gamma v_H/c$. We still consider $(r_V, \Delta_r, \delta_V) = (3.5 \, \mathrm{Gpc}, 0.42 \, \mathrm{Gpc}, -0.058)$ with a direction $(l,b)=(240^\circ,33^\circ)$ that satisfies the requirements of cosmic dipole directions. The required quasar dipole at different redshifts seen by a 219 Mpc observer can now be shown as in Fig.~\ref{p2_dpqua_z}. We can tell that the overall dipole induced by our void model is still able to meet the quasar dipole requirement well at the corresponding redshift.

In previous work simulations had been done to show how the amplitude and direction of the dipole of CatWISE quasar sample \cite{Eisenhardt_2020} and the kinematic interpretation of the CMB dipole differs from each other \cite{Secrest:2020has}. In this work, we run the same simulations not only for the kinematic interpretation of the CMB dipole, but also for our void model with reconstructed peculiar velocity. Note that the quasar dipole of the CatWISE sample used in the simulations is $|\boldsymbol{\mathcal{D}}_\text{quasar}|=1.554\times10^{-2}, (l,b)=(238\dotdeg2,28\dotdeg8)$, slightly different from the median value of our previous quasar dipole due to a minor change of the flux density cut \cite{Secrest:2022uvx}. For this case we change the position to the center of the void to $240$ Mpc for a better fit of this dipole while maintaining the CMB dipole in constraint. For the peculiar velocity observed with large uncertainty, we simply assume the median value $|\mathbf{v}_s|=278 \, \text{km}\cdot\text{s}^{-1}, (l,b)=(309^\circ,22^\circ)$. In the MC simulations for the no void case, the velocity of the observer is set to be the velocity directly inferred from the CMB dipole. When we switch to the void case, this velocity is then the median of the combined distribution of observed peculiar velocity $\mathbf{v}_H$ and the velocity from the void pull $\mathbf{v}_\text{s}$. After obtaining the simulated dipole distribution based on this combined velocity, we add the dipole caused by the void structure $\boldsymbol{\mathcal{D}}_1$ and have the final simulated dipole distributions. 

\begin{figure}[h]
    \centering
    \begin{subfigure}{0.52\textwidth}
        \includegraphics[width=\linewidth]{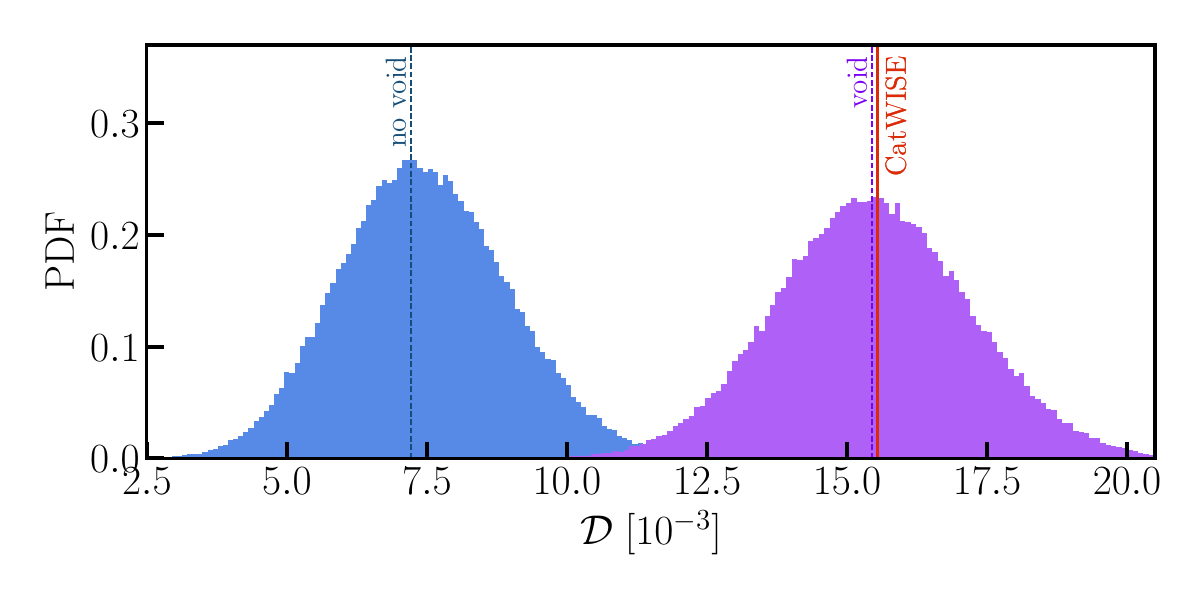}
    \end{subfigure}%
    \begin{subfigure}{0.4\textwidth}
        \includegraphics[width=\linewidth]{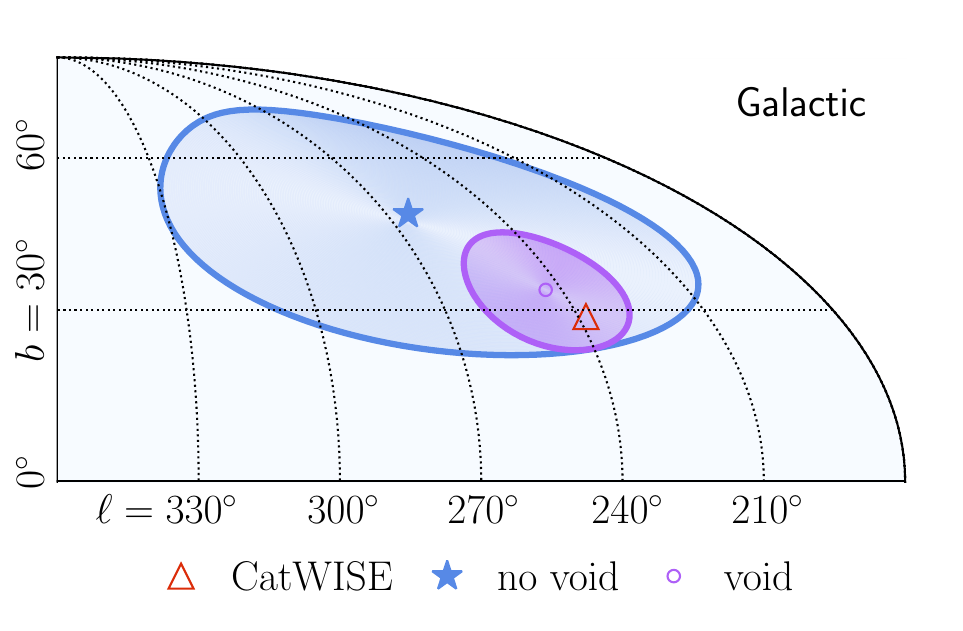}
    \end{subfigure}
    \caption{\textit{Left panel:} Solid vertical line is the dipole amplitude in the CatWISE quasar sample $|\boldsymbol{\mathcal{D}}_\text{quasar}|$. The blue plot is the distribution of the simulated dipoles calculated in a standard $\Lambda$CDM model with no void and the velocity of the observer set to be the kinematic interpretation of the CMB dipole. The purple plot is the distribution of simulated dipoles in our void model with the median value of the peculiar velocity observed for the solar motion. The dashed vertical lines are the median value of the corresponding distributions. \textit{Right panel:} Dipole directions $(l,b)$  of both simulated distributions in Galactic coordinates with uncertainty region ($2\sigma$). The direction of the dipole in the CatWISE quasar sample and the center directions of the simulated dipole distributions are marked with the corresponding legends.}
    \label{p2_mc}
\end{figure}

The results of the simulations of the final observed dipole $\boldsymbol{\mathcal{D}}_{\text{quasar}}$ are shown in Fig.~\ref{p2_mc}. Comparing what a $\Lambda$CDM observer would see (blue plots) with what an off-center gigaparsec void observer would see (purple plots), we see that in this simulation, the dipole amplitude results given by the void model are much closer to the dipole amplitude of the CatWISE quasar sample than that given by the purely kinematic interpretation of the CMB dipole in standard $\Lambda$CDM cosmology. The center direction of the dipole results given by the void model is also closer to the dipole direction of the CatWISE quasar sample, allowing its corresponding distribution to include the CatWISE direction even when its uncertainty region becomes smaller due to the larger dipole amplitude. This shows that the void model can offer a good explanation to the dipole of the observed quasar samples.

\section{Conclusions and Remarks}
To summarize, we propose that the existence of a Gpc-scale local void could affect the validity of the cosmological principle, and reconciling the related dipolar tension. In particular, we study the induced CMB dipole and quasar dipole in the view of an off-center observer inside a Gpc-scale local void. Due to the intrinsic matter distribution anisotropy in an off-center void, the detected photons from different directions have experienced distinct cosmic expansion histories, which causes an anisotropy in their redshift. Such a redshift anisotropy could induce cosmic dipoles by affecting the CMB temperature fluctuation and the analysis in quasar number counting. Also, the intrinsic matter anisotropy in the view of an off-center observer can induce an angular matter density dipole, which should be an approximation of the quasar number dipole. In this Gpc-scale local void scenario, the induced dipole structures in the view of an off-center observer are distinct at different redshifts, and a peak feature in dipole amplitude appears around void boundary. When the void boundary is located at $z = 1.2$, such a peak feature in dipole amplitude can help reconcile the dipolar tension between CMB dipole and quasar dipole.

In our analysis, three dipole effects are taken into considerations those are the dipole caused by void structure, void-induced peculiar velocity, and the Local Group peculiar velocity. We then consider two benchmark void models to evaluate induced dipoles. The first benchmark model considers the contribution from void structure and void-induced peculiar velocity, a local void with void profile $(r_V, \Delta_r, \delta_V) = (3.4 \, \mathrm{Gpc}, 0.41 \, \mathrm{Gpc}, -0.05)$, an off-center observer with $292 \, \mathrm{Mpc}$ away from the void center and relative orientation $(l,b) = $ (264\textdegree, 48\textdegree) in galactic coordinates, could observe the CMB dipole $\mathcal{D}_\mathrm{CMB} \simeq 1.23 \times 10^{-3}$ and quasar dipole $\mathcal{D}_\mathrm{quasar} \simeq 1.48 \times 10^{-2}$. The second benchmark model considers the contribution from all three dipole effects, a local void with void profile $(r_V, \Delta_r, \delta_V) = (3.5 \, \mathrm{Gpc}, 0.42 \, \mathrm{Gpc}, -0.058)$, an off-center observer with $219 \, \mathrm{Mpc}$ away from void center and relative orientation $(l,b) = $ (240\textdegree, 33\textdegree) in galactic coordinates, can observe a CMB dipole $\mathcal{D}_c \simeq 1.07 \times 10^{-3}$ and quasar dipole $\mathcal{D}_q \simeq 1.41 \times 10^{-2}$, and adding the contribution from a reconstructed peculiar velocity $|\mathbf{v}_s|=278\pm352\text{km}\cdot\text{s}^{-1}$ in the direction of $(l,b)=(309^\circ\pm79^\circ,22^\circ\pm80^\circ)$, the dipole in this benchmark model matches the observed CMB and quasar dipoles. Meanwhile, we use the CatWISE quasar sample to test the dipolar tension performance in local void scenario, where void profile is the same as the second benchmark model with an off-center position $240 \, \mathrm{Mpc}$ and we fix the peculiar velocity as $|\mathbf{v}_s|=278 \, \text{km}\cdot\text{s}^{-1}, (l,b)=(309^\circ,22^\circ)$. The MC simulation shows that the dipolar tension between the amplitude of CMB dipole and quasar dipole decreases from $4.9 \sigma$ to within $1 \sigma$, and the direction of quasar dipole in benchmark model is within $2 \sigma$ range of observed quasar dipole as shown in Fig.~\ref{p2_mc}.

In this work, motivated by the dipolar tension, we mainly focus on comparing the cosmic dipoles induced by an off-center observer in a local void. In this scenario, the void profile and the Local Group peculiar velocity would significantly impact our result. To determine a suitable void profile, an accurate LG peculiar velocity should be obtained. However current uncertainty in LG peculiar velocity that reconstructed from Cosmicflow-4 data is very large \cite{Tully:2022rbj, Courtois:2022mxo, Stiskalek:2025xaw}, and causes the difficulty in determining the void profile. Our work highlights the need for further investigations in LG peculiar velocity to improve understanding and for more accurate predictions in void profiles. In addition, the shape of the void would also change our result, if the local void is not spherical, additional anisotropies of higher multipoles may also be introduced, and affects the our analysis in induced dipole.

In our calculation, we use the void profiles $(r_V, \Delta_r, \delta_V) = (3.4 \, \mathrm{Gpc}, 0.41 \, \mathrm{Gpc}, -0.05)$ and $(3.5 \, \mathrm{Gpc}, 0.42 \, \mathrm{Gpc}, -0.058)$ as benchmark models, which is consistent with observed CMB dipole and quasar number dipole. However, such a void profile cannot fully explain some other observations such as Hubble tension as discussed above. In this case, an overall picture of void profile should take all possible effects into consideration. For instance, as an off-center observer in a local void, we can not only observe the Hubble tension and dipolar tension, but also find the existence of $S_8$ tension in observation \cite{Lee:2013uzp, Ichiki:2015gia}. After considering all possible effects including Hubble tension, dipolar tension and $S_8$ tension, a physical parameter of this void profile can be determined, which we leave for future study. 

It is also important to note that due to the off-center observer in a void, the dipolar anisotropy should consistently exist in many cosmic signals, e.g., Type Ia SNe \cite{Alnes:2006uk, Sun:2018cha, Mohayaee:2020wxf, Krishnan:2021jmh, Aluri:2022hzs}, large scale structure \cite{Globus:2018svy, Migkas:2021zdo}, $21 \mathrm{cm}$ background \cite{Cooray:2006km} and gravitational wave background \cite{NANOGrav:2020bcs, NANOGrav:2023gor}. If we are indeed located in a void, the void profile can be understood better by a combined study of these signals.

\section*{Acknowledgment} 
We would like to thank the anonymous referee for the insightful suggestions that helped us greatly improve this work. We also thank Haipeng An and Zhong-Zhi Xianyu for helpful discussion. This work is supported in part by the National Key R\&D Program of China (2021YFC2203100),and the GRF grant 16303621 and the RFS grant RFS2425-6S02 by the RGC of Hong Kong SAR.

\appendix

\section{Simulation of combined vector distribution\label{app}}

For two vector distributions in spherical coordinates $(\bar{v}_1\pm u_{v1},\bar{l}_1\pm u_{l1},\bar{b}_1\pm u_{b1}), (\bar{v}_2\pm u_{v2},\bar{l}_2\pm u_{l2},\bar{b}_2\pm u_{b2})$, each $(v,l,b)$ can be converted into the corresponding Cartesian coordinates $(x,y,z)$
\begin{align}
    x&=v\cos l\cos b,\\
    y&=v\sin l\cos b,\\
    z&=v\sin b.
\end{align}
If we set the number of simulations to be a large number n, then a list of sample Cartesian coordinates in the combined vector distributions can be obtained by directly summing the Cartesian coordinates generated and converted from the distributions in spherical coordinates. Here we use non-relativistic addition and assume uncorrelated errors
\begin{equation}
    (x_i,y_i,z_i)=(x_{1i},y_{1i},z_{1i})+(x_{2i},y_{2i},z_{2i})\ \ \ \ i=1,2,...n.
\end{equation}
While the mean vector in spherical coordinates $(\bar v,\bar l,\bar b)$ can be directly derived from $(\bar x,\bar y,\bar z)=\sum^n_{i=1}(x_i,y_i,z_i)/n$, the corresponding uncertainty follows the law of propagation
\begin{align}
    u_v&=\sqrt{\left(\frac{\partial v}{\partial x}\right)^2u_x^2+\left(\frac{\partial v}{\partial y}\right)^2u_y^2+\left(\frac{\partial v}{\partial z}\right)^2u_z^2},\\
    u_l&=\sqrt{\left(\frac{\partial l}{\partial x}\right)^2u_x^2+\left(\frac{\partial l}{\partial y}\right)^2u_y^2+\left(\frac{\partial l}{\partial z}\right)^2u_z^2},\\
    u_b&=\sqrt{\left(\frac{\partial b}{\partial x}\right)^2u_x^2+\left(\frac{\partial b}{\partial y}\right)^2u_y^2+\left(\frac{\partial b}{\partial z}\right)^2u_z^2}
\end{align}
where $u_x=\sqrt{\frac{1}{n-1}\sum^n_{i=1}(x-x_i)^2}$ is the uncertainty of coordinate $x$ and the same for $y,z$. The result of the combined vector distribution is then $(\bar v\pm u_v, \bar l\pm u_l, \bar b\pm u_b)$.

\end{document}